\documentclass[aps,prx,twocolumn,superscriptaddress]{revtex4-2}

\usepackage{bookmark}   
\usepackage[UKenglish]{babel}
\usepackage{colorprofiles}
\usepackage[doipre={doi:}]{uri}   
\hypersetup{hidelinks}

\usepackage{csquotes}

\usepackage{import}

\usepackage{etoolbox}
\apptocmd{\sloppy}{\hbadness 10000\relax}{}{} 

\usepackage{mathtools}
\usepackage{physics}
\usepackage{amsfonts}
\usepackage{mathdots}   
\usepackage{mathrsfs}   
\usepackage{mdframed}
\usepackage{bigints}    
\usepackage{xfrac}      
\usepackage{stmaryrd}   
\SetSymbolFont{stmry}{bold}{U}{stmry}{m}{n}   

\usepackage{amsthm}
\usepackage{thmtools}
\usepackage{thm-restate}

\newtheorem{lemma}{Lemma}

\newtheorem{definition}{Definition}
\newtheorem{proposition}{Proposition}

\usepackage{capt-of}
\usepackage{placeins}

\usepackage{bm}   

\usepackage{blkarray}   

\usepackage{graphicx}
\graphicspath{../figures}
\usepackage{tikz}
\usepackage{tikzfig}
\usetikzlibrary{external}
\usepackage{pgfplots} 
\pgfplotsset{compat=newest, compat/show suggested version=false} 

\tikzstyle{box}=[draw=black, shape=rectangle, fill=white, minimum size=.95em, inner sep=0.15em, scale=0.85, font={\scriptsize}]
\tikzstyle{gbox}=[box, draw=black, shape=rectangle, fill={zx_green}, tikzit fill={rgb,255: red,181; green,215; blue,181}, tikzit shape=rectangle]
\tikzstyle{hbox}=[box, draw=black, shape=rectangle, fill=yellow, minimum size=.55em]
\tikzstyle{gn}=[draw=black, shape=circle, fill={zx_green}, inner sep=0.7mm, minimum width=0pt, minimum height=0pt, tikzit fill={rgb,255: red,181; green,215; blue,181}]
\tikzstyle{gn_phase}=[shape=rectangle, fill={zx_green}, draw=black, minimum size=1.2em, rounded corners=0.5em, inner sep=0.2em, outer sep=-0.2em, scale=0.8, font={\scriptsize\boldmath}, tikzit shape=circle, tikzit fill={rgb,255: red,130; green,188; blue,130}]
\tikzstyle{rn}=[gn, fill={zx_red}, draw=black, tikzit fill={rgb,255: red,255; green,165; blue,165}]
\tikzstyle{rn_phase}=[{gn_phase}, fill={zx_red}, draw=black, tikzit fill={rgb,255: red,215; green,96; blue,96}, font={\scriptsize\boldmath}]
\tikzstyle{pn}=[gn, fill={zx_pink}, draw=black, tikzit fill=pink]
\tikzstyle{pn_phase}=[{gn_phase}, fill={zx_pink}, draw=black, tikzit fill=red]
\tikzstyle{rtriang}=[shape=isosceles triangle, fill=yellow, draw=black, isosceles triangle stretches=true, inner sep=0.8pt, minimum width=0.25cm, minimum height=2mm]
\tikzstyle{ltriang}=[rtriang, shape=isosceles triangle, fill=yellow, draw=black, shape border rotate=180]
\tikzstyle{utriang}=[rtriang, shape=isosceles triangle, fill=yellow, draw=black, shape border rotate=90]
\tikzstyle{dtriang}=[rtriang, shape=isosceles triangle, fill=yellow, draw=black, shape border rotate=-90]
\tikzstyle{lmat}=[shape=signal, signal to=west, signal from=east, fill={zx_grey}, draw=black, minimum height=6pt, inner sep=.75pt, font={\scriptsize \boldmath}, tikzit fill=gray, tikzit category=GLA]
\tikzstyle{rmat}=[lmat, shape=signal, signal to=east, signal from=west, tikzit fill=gray, tikzit category=GLA]
\tikzstyle{dmat}=[lmat, shape=signal, signal to=west, signal from=east, tikzit fill=gray, tikzit category=GLA, rotate=90]
\tikzstyle{umat}=[lmat, shape=signal, signal to=east, signal from=west, tikzit fill=gray, tikzit category=GLA, rotate=90]
\tikzstyle{uw}=[shape=isosceles triangle, isosceles triangle stretches=true, fill=black, draw=black, minimum width=0.4cm, minimum height=3mm, inner sep=1pt, shape border rotate=90]
\tikzstyle{dw}=[shape=isosceles triangle, isosceles triangle stretches=true, fill=black, draw=black, minimum width=0.4cm, minimum height=3mm, inner sep=1pt, shape border rotate=-90]
\tikzstyle{lw}=[shape=isosceles triangle, isosceles triangle stretches=true, fill=black, draw=black, minimum width=0.4cm, minimum height=3mm, inner sep=1pt, shape border rotate=180]
\tikzstyle{rw}=[shape=isosceles triangle, isosceles triangle stretches=true, fill=black, draw=black, minimum width=0.4cm, minimum height=3mm, inner sep=1pt]
\tikzstyle{d_split}=[shape=trapezium, fill=white, draw=black, inner sep=0pt, trapezium stretches body, text width=15pt, text height=7pt]
\tikzstyle{d_merge}=[{d_split}, shape=trapezium, draw=black, rotate=180]
\tikzstyle{sd_split}=[shape=trapezium, fill=white, draw=black, inner sep=0pt, trapezium stretches body, text width=10pt, text height=5pt]
\tikzstyle{sd_merge}=[{sd_split}, shape=trapezium, draw=black, rotate=180]
\tikzstyle{wire label}=[font={\tiny}, auto]
\tikzstyle{control}=[draw=black, shape=circle, fill=black, inner sep=0.5mm]

\tikzstyle{symmetriser}=[-, very thick]
\tikzstyle{braceedge}=[-, decorate, decoration={brace, amplitude=2mm, raise=-1mm}]
\tikzstyle{dotsedge}=[-, dotted, decoration={brace, amplitude=2mm, raise=-1mm}]
\tikzstyle{filllayer}=[-, fill={rgb,255: red,178; green,227; blue,247}, opacity=0.5]

\definecolor{zx_grey}{RGB}{211,211,211}
\definecolor{zx_red}{RGB}{232,165,165}
\definecolor{zx_pink}{RGB}{255, 130, 160}
\definecolor{zx_green}{RGB}{216,248,216}

\usepackage{xtab}
\usepackage{tabu}

\usepackage[capitalise, noabbrev]{cleveref}

\usepackage[percent]{overpic} 

\ifdef{\C}
  {\renewcommand{\C}{\mathbb{C}}}
  {\newcommand{\C}{\mathbb{C}}}
\newcommand{\minu}{\texttt{-}}
\newcommand{\plus}{\texttt{+}}
\newcommand{\ccdots}{{\cdot}\mkern1mu{\cdot}\mkern1mu{\cdot}}

\newcommand{\interp}[1]{\left\llbracket#1\right\rrbracket}

\newcommand{\bR}{\begin{color}{red}}
\newcommand{\bB}{\begin{color}{blue}}
\newcommand{\bM}{\begin{color}{magenta}}
\newcommand{\bC}{\begin{color}{cyan}}
\newcommand{\bW}{\begin{color}{white}}
\newcommand{\bBl}{\begin{color}{black}}
\newcommand{\bG}{\begin{color}{green}}
\newcommand{\bY}{\begin{color}{yellow}}
\newcommand{\bBr}{\begin{color}{brown}}
\newcommand{\e}{\end{color}}

\newcommand{\Wthree}[6]{\begin{pmatrix} #1 & #2 & #3 \\ #4 & #5 & #6 \end{pmatrix}}
\newcommand{\Wfour}[9]{\begin{pmatrix} #1 & #2 & #3 & #4 \\ #5 & #6 & #7 & #8 \end{pmatrix}^{(#9)}}

\newcommand{\tikzrefsize}[1]{\scriptsize{#1}}
\newcommand{\axref}[1]{\tikzrefsize{\eqref{rule:#1}}}
\newcommand{\lemref}[1]{\tikzrefsize{\eqref{#1}}}

\newcommand{\Mod}[1]{\ (\mathrm{mod}\ #1)}

\begin{document}

\title{\normalsize Beyond Penrose tensor diagrams with the ZX calculus: Applications to quantum computing, quantum machine learning, condensed matter physics, and quantum gravity}

\author{Quanlong Wang}
\affiliation{%
 Quantinuum, 17 Beaumont Street, Oxford, OX1 2NA, United Kingdom
}%

\author{Richard D.~P.~East}%
\affiliation{%
 Haiqu Inc., San Francisco, CA 94103, USA
}%

\author{Razin A.~Shaikh}%
\affiliation{%
 Quantinuum, 17 Beaumont Street, Oxford, OX1 2NA, United Kingdom
}%
\affiliation{Department of Computer Science, University of Oxford, OX1 3QD, United Kingdom}

\author{Lia Yeh}%
\affiliation{%
 Department of Computer Science and Technology, University of Cambridge, CB3 0FD, United Kingdom
}
\altaffiliation[This work was done while LY was at ]{Quantinuum and at the Department of Computer Science, University of Oxford.}

\author{Boldizsár Poór}%
\affiliation{%
 Quantinuum, 17 Beaumont Street, Oxford, OX1 2NA, United Kingdom
}%
\affiliation{Department of Computer Science, University of Oxford, OX1 3QD, United Kingdom}

\author{Bob Coecke}%
\affiliation{%
 Relational Intelligence, Oxford, OX1 5PE, United Kingdom
}%
\affiliation{Perimeter Institute for Theoretical Physics, Waterloo, ON N2L 2Y5, Canada}
\altaffiliation[This work was done while BC was at ]{Quantinuum}

\date{\today}

\begin{abstract}
  We introduce the \emph{Spin-ZX calculus} as an elevation of Penrose's diagrams and associated binor calculus to the level of a formal diagrammatic language.
The power of doing so is illustrated by the variety of scientific areas we apply it to:
permutational quantum computing, quantum machine learning, condensed matter physics, and quantum gravity.
Respectively, we analyse permutational computing transition amplitudes, evaluate barren plateaus for $SU(2)$ symmetric ans\"atze, study properties of AKLT states, and derive the minimum quantised volume in loop quantum gravity.

Our starting point is the mixed-dimensional ZX calculus, a purely diagrammatic language that has been proven to be complete for finite-dimensional Hilbert spaces.
That is, any equation that can be derived in the Hilbert space formalism, can also be derived in the mixed-dimensional ZX calculus.
We embed the Spin-ZX calculus inside the mixed-dimensional ZX calculus, rendering it a quantum information flavoured diagrammatic language for the quantum theory of angular momentum, i.e.\@ $SU(2)$ representation theory.
We diagrammatically derive the fundamental spin coupling objects --- such as Clebsch-Gordan coefficients, symmetrising mappings between qubits and spin spaces, and spin Hamiltonians --- under this embedding.

Our results establish the Spin-ZX calculus as a powerful tool for representing and computing with \( \mathrm{SU}(2) \) systems graphically, offering new insights into foundational relationships and paving the way for new diagrammatic algorithms for theoretical physics.
\end{abstract}
\maketitle

\section{Introduction}\label{sec:introduction}


The quantum theory of angular momentum,  familiar to mathematicians as the representation theory of $SU(2)$ and its Lie algebra, studies the quantum property of spins and their coupling.
It is of great importance in a broad range of research areas such as quantum chemistry,  condensed matter physics, particle physics, quantum computing and quantum gravity. 
However, calculations involving the coupling of spins quickly become unwieldy using algebraic methods.
This has led to the development of diagrammatic methods to represent and evaluate complex expressions --- containing symmetrisers and Clebsch-Gordan coefficients --- in an intuitive way.

Graphical tensor notation was introduced by Roger Penrose 
mainly as a calculational tool for areas of physics that work with multilinear functions or tensors~\cite{penroseApplicationsNegativeDimensional1971}.
While Yutsis was studying quantum chemistry, he developed a graphical notation for SU(2), i.e.\@ spin, to aid his calculations.
These two roots later found a unification in Penrose's binor calculus --- a graphical tenors network notation designed for SU(2).
Due to their specificity, they remained a niche technical tools for many years.

Inspired by the works of Penrose, new diagrammatic languages were also being developed in computer science to represent information flows in programming language semantics~\cite{huetConfluentReductions1980,girardLinearLogic1987}.
Their mathematical status was elucidated in the 90's by Joyal and Street~\cite{joyalGeometryTensor1991}.
To unify these, in around 2004, Categorical Quantum Mechanics (CQM) aimed to provide a new formalism for quantum theory~\cite{abramskyCategoricalSemanticsQuantum2004}.
This led to the development of the ZX calculus~\cite{coeckeInteractingQuantumObservables2008}, a specialisation of CQM, which is now widely used in quantum information and quantum computing.

ZX calculus~\cite{coeckeInteractingQuantumObservables2008,coeckeInteractingQuantumObservables2011} offers a diagrammatic approach to reasoning about qubit quantum systems.
Any quantum computation between qubits, or more generally, any linear map between qubits, can be represented as a ZX-diagram, a type of tensor network created from a small number of generating elements.
The ZX calculus is a sound, universal, and complete language for qubit linear algebra~\cite{ngUniversalCompletionZXcalculus2017,vilmartNearMinimalAxiomatisationZXCalculus2019}.
This means that any proof it derives is correct, it can represent any linear map, and we can prove any equality between linear maps using only the rules of the calculus, making it as powerful as the Hilbert space formalism.
Recent works have extended the ZX calculus beyond qubits to higher dimensional quantum systems in both finite~\cite{poorQupitStabiliserZXtravaganza2023,poorCompletenessArbitraryFinite2023,wangCompletenessQufiniteZXW2024,poorZXcalculusComplete2024} and infinite~\cite{shaikhFockedupZX2024,nagayoshiZXGraphical2024,boothCompleteEquational2024} dimensional Hilbert spaces.

This approach to quantum mechanics is philosophically different from the usual physicists' analysis of quantum systems that has come before;
it is born from the theoretical computer science perspective focusing on processes and their compositions, specifically formalised through applied category theory~\cite{abramskyCategoricalQuantumMechanics2008}.
The calculus has found practical use in a number of areas across quantum information, with results in measurement-based quantum computation~\cite{debeaudrapCircuitExtractionZXDiagrams2022,duncanRewritingMeasurementBasedQuantum2010,kissingerUniversalMBQCGeneralised2019}, topological quantum computation~\cite{debeaudrapPauliFusionComputational2020,debeaudrapZXCalculusLanguage2020,gidneyEfficientMagicState2019,hanksEffectiveCompressionQuantum2020}, quantum error correction~\cite{tanSATScalpel2024,chancellorGraphicalStructuresDesign2023,huangGraphicalCSSCode2023,kissingerPhasefreeZXDiagrams2022,rodatzFloquetifyingStabiliser2024,rodatzFaultTolerance2025}, quantum chemistry~\cite{defeliceLightMatterInteractionZXW2023,shaikhHowSum2023}, and quantum circuit compilation and optimisation~\cite{debeaudrapFastEffective2020,cowtanPhaseGadgetSynthesis2020,duncanGraphtheoreticSimplificationQuantum2020,kissingerReducingTcountZXcalculus2020}.


In this paper we bring together Penrose's binor calculus and ZX calculus, presenting a formal diagrammatic language for the analysis of spin.
The extension of diagrammatic language techniques from qubits to spin has been investigated previously~\cite{eastPhysiqueFormelleSpin2022,eastSpinnetworksZXcalculus2022,eastAKLTStatesZXDiagramsDiagrammatic2022}.
This paper can be seen as a completion of the work initiated there to provide a compositional framework for quantum algorithms beyond qubits, through the development of a calculus specific to physical symmetries, largely spin systems. 
The underlying aim of this research program is to create a bridge between quantum algorithms based upon qubits and qudits as an engineering resource, and algorithms based on the representation theory of spin as a fundamental physical symmetry.
The concrete medium term ambition is the construction of a formal diagrammatic language for the full standard model.
Given the model itself is founded on the representations of the three groups $U(1)\cross SU(2)\cross SU(3)$, this paper amounts to a formal diagrammatic language for the second of these. 


\subsection{Whistle-stop pictorial tour}


Let us first give a details-light diagram-heavy guide to the paper that will be fleshed out in the main text. In this work, we introduce the Spin-ZX calculus — a diagrammatic formalism that unifies SU(2) recoupling theory and quantum informational methods by embedding Penrose’s classical diagrams as a fragment of the mixed-dimensional ZX calculus. Mixed-dimensional ZX is made up of elements:

\begin{itemize}

	\item \emph{Z spider},
	\[
	\input{qufit-generators/mixed-zbox.tikz}
	\quad \overset{\interp{\cdot}}{\longmapsto} \quad
	\sum_{j=0}^{\min{\{d_i\}_i} - 1} a_j
	\ket{j, \cdots\!, j} \bra{j, \cdots\!, j}
	\]
	where $a_0 \coloneqq 1$ and $\overrightarrow{a} = (a_1, \cdots, a_{\min{\{d_i\}_i}-1})$.

	\item \emph{X spider},
	\[
	\input{qufit-generators/quditrspiderclassicnm.tikz}
	\quad \overset{\interp{\cdot}}{\longmapsto}
	\sum_{
		\substack{
			i_1+\cdots\!+ i_m + j\\
			\equiv j_1+\cdots\!+j_n \! \Mod{d}
		}
	}
	\ket{i_1, \cdots\!, i_m}\bra{j_1, \cdots\!, j_n}
	\]
	where  $j \in \mathbb{Z}$.
	Two X spiders labelled $j$ and $j'$ are equal by definition if $j \equiv j' \! \Mod{d}$.
\end{itemize}

We use these to construct the generators of Spin-ZX calculus which, as explained further below, are composed of Yutsis diagrams~\cite{yutsisMathematicalApparatusTheory1962} which can be seen as spin-invariant components that can be composed to form Penrose's spin networks~\cite{penroseApplicationsNegativeDimensional1971}.

 \begin{figure*}[ht]
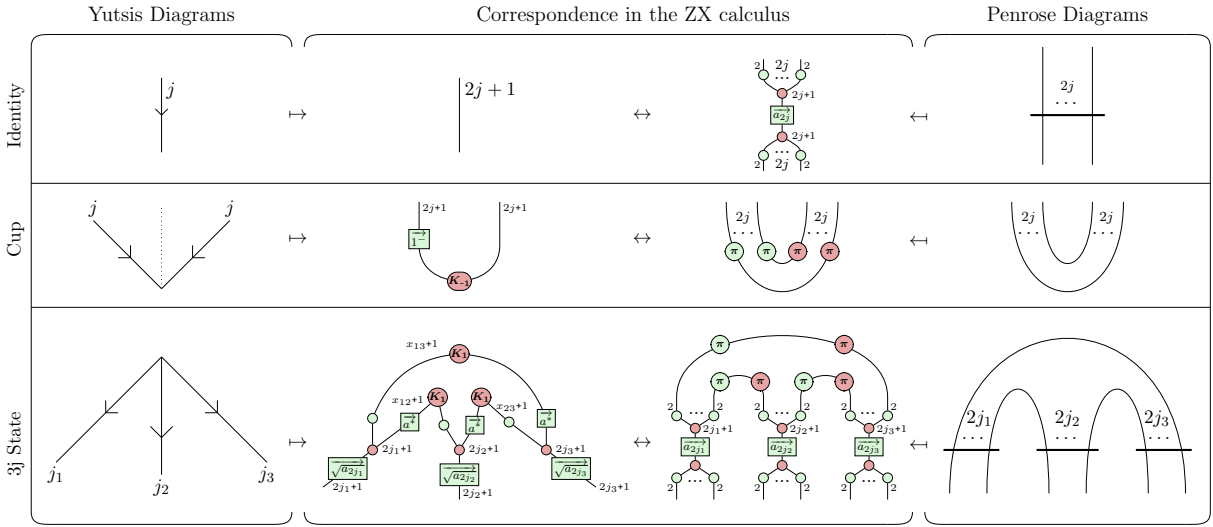

	\centering
	\ctikzfigscale{0.66}{figures/intro-table}
	\caption{Generators of the Spin-ZX calculus. Here, $j, j_1, j_2, j_3 \in \mathbb{N}/2$, $x_{lm} = j_l + j_m - j_n$, where $\protect\overrightarrow{a_n} = \left( \frac{1}{\binom{n}{1}}, \cdots, \frac{1}{\binom{n}{k}}, \cdots, \frac{1}{\binom{n}{n}} \right)$, $\protect\overrightarrow{1^{-}} = \left( (-1)^1, \cdots, (-1)^{2j} \right)$, and $\protect\overrightarrow{\frac{1}{a^{-}}} = \left((-1)^1\binom{x_{k \ell}}{1}, \cdots, (-1)^{x_{k \ell}}\binom{x_{k \ell}}{x_{k \ell}}\right)$.}
	\label{fig:generators-intro}
\end{figure*}

In \cref{fig:generators-intro} , we can see the following:
\begin{itemize}
	\item The \emph{Yutsis diagram} (on the left);
	\item The \emph{Penrose diagram} (on the right); and
	\item The \emph{ZX diagrams} (in the middle), which are their translations into our modern diagrammatic language.
\end{itemize}
In the left half of \cref{fig:generators-intro}, each row has the following \emph{interpretation} in terms of standard SU(2) representation theory:
\begin{itemize}
	\item The \emph{identity} operator on the spin-$j$ space is $\displaystyle \sum_{m} \ket{j;m}\bra{j;m}$;
	\item The \emph{cup}, which represents the map from an input to an output wire, or a spin-$j$ Hilbert space to its dual, is $\displaystyle \sqrt{2j+1}\sum_{m} (-1)^{j+m+1} \bra{j;m, \, j;-m}$; and
	\item The \emph{3j--state}, which is equivalent~\footnote{This equivalence is up to the 3j--state being symmetric with respect to the spins. Rather than a map from two of the spins to a third, it represents how all three spins combine to spin-0.} to the Clebsch-Gordan map that tells us how two spins combine, is $\displaystyle \sum_{m_1,m_2,m_3} \mqty(j_1 & j_2 & j_3 \\ m_1 & m_2 & m_3) \ket{j_1;m_1,\, j_2;m_2,\, j_3;m_3}$.
\end{itemize}

The ZX diagrams faithfully satisfy the correspondence between the left and right halves of \cref{fig:generators-intro}, by natively representing the following isometry map to the symmetrised subspace of $(\mathbb{C}^2)^{\otimes n}$ which defines this correspondence:
\[
	\input{figures/binor/isometry-def.tikz}
\]
where
\[\overrightarrow{\sqrt{a_n}} = \left( \frac{1}{\sqrt{\binom{n}{1}}}, \cdots, \frac{1}{\sqrt{\binom{n}{k}}}, \cdots, \frac{1}{\sqrt{\binom{n}{n}}} \right)
\]

Composing this isometry and its adjoint yields the projection onto the symmetrised subspace, called the \emph{symmetriser}:
\[
\scalebox{1}{\input{figures/binor/sym-j.tikz}}
\qquad\longmapsto\quad
\scalebox{1}{\input{figures/qufit-applications/spinjiireps2v2.tikz}}
\]
where
\[\overrightarrow{a_{2j}} = \left( \frac{1}{\binom{2j}{1}}, \cdots, \frac{1}{\binom{2j}{k}}, \cdots, \frac{1}{\binom{2j}{2j}} \right).
\]
On the left, we can see Penrose's notation for the symmetriser, which builds a spin-$j$ space by  symmetrising the $2j$ spin-$\frac{1}{2}$ wires.
On the right, we have the ZX calculus representation of the symmetriser.
This enables us to exactly represent the decomposition of higher spins into the symmetric subspace of multiple copies of spin-$\frac{1}{2}$.

As a diagrammatic calculus, it also comes with a collection of rewrite rules given in \cref{Table:Spin-ZX-rewrites}.

With the calculus outlined we apply it in a number of different settings. In quantum computing theory we tackle the permutational quantum computing (PQC) model:
\[
\scalebox{0.75}{\begin{tikzpicture}
	\begin{pgfonlayer}{nodelayer}
		\node [style=none] (0) at (-7.25, 5.25) {$\frac{1}{\sqrt{3}}$};
		\node [style=none] (1) at (-6.175, 5.25) {};
		\node [style=none] (2) at (-1.175, 3.5) {};
		\node [style=none] (3) at (-1.175, 7) {};
		\node [style=none] (4) at (-4.925, 3.5) {};
		\node [style=none] (5) at (-4.925, 7) {};
		\node [style=none] (6) at (-2.675, 4.625) {};
		\node [style=none] (7) at (-3.675, 5.25) {};
		\node [style=none] (8) at (-4.675, 5.25) {};
		\node [style=wire label] (9) at (-3.3, 4.625) {$1$};
		\node [style=wire label] (10) at (-5.175, 5.25) {$\frac{1}{2}$};
		\node [style=none] (11) at (-1.675, 4) {};
		\node [style=none] (12) at (-1.675, 5.25) {};
		\node [style=none] (13) at (-1.65, 6.5) {};
		\node [style=none] (14) at (-0.675, 6.5) {};
		\node [style=none] (15) at (1.325, 6.5) {};
		\node [style=none] (16) at (-0.675, 5.25) {};
		\node [style=none] (17) at (1.325, 5.25) {};
		\node [style=none] (18) at (-0.675, 4) {};
		\node [style=none] (19) at (1.325, 4) {};
		\node [style=none] (20) at (6.825, 5.25) {};
		\node [style=none] (21) at (1.825, 3.5) {};
		\node [style=none] (22) at (1.825, 7) {};
		\node [style=none] (23) at (5.575, 3.5) {};
		\node [style=none] (24) at (5.575, 7) {};
		\node [style=none] (25) at (3.325, 4.625) {};
		\node [style=none] (26) at (4.325, 5.25) {};
		\node [style=none] (27) at (5.325, 5.25) {};
		\node [style=wire label] (28) at (3.95, 4.625) {$0$};
		\node [style=wire label] (29) at (5.825, 5.25) {$\frac{1}{2}$};
		\node [style=none] (30) at (2.325, 4) {};
		\node [style=none] (31) at (2.325, 5.25) {};
		\node [style=none] (32) at (2.3, 6.5) {};
		\node [style=none] (33) at (-7.175, 0) {$=$};
		\node [style=none] (34) at (-5.675, 0) {$\frac{1}{\sqrt{3}}$};
		\node [style=gn] (35) at (-2.275, -1.5) {};
		\node [style=rn] (36) at (-3.925, -1.5) {};
		\node [style=gbox] (37) at (1.225, -0.75) {$\overrightarrow{a}$};
		\node [style=rn] (38) at (2.225, -0.75) {};
		\node [style=none] (39) at (3.1, 0) {};
		\node [style=none] (40) at (3.1, -1.5) {};
		\node [style=rn] (41) at (0.075, -0.75) {};
		\node [style=none] (42) at (-0.8, 0) {};
		\node [style=none] (43) at (-0.8, -1.5) {};
		\node [style=wire label] (44) at (7, -1.775) {$2$};
		\node [style=wire label] (45) at (3, 0.275) {$3$};
		\node [style={gn_phase}] (46) at (-3.525, 1.5) {$\pi$};
		\node [style={rn_phase}] (47) at (-3.525, 0) {$\pi$};
		\node [style=rn] (48) at (8.275, 1.5) {};
		\node [style=wire label] (49) at (7.075, 1.725) {$2$};
		\node [style={gn_phase}] (50) at (7.875, 0) {$\pi$};
		\node [style={rn_phase}] (51) at (7.875, -1.5) {$\pi$};
		\node [style=wire label] (52) at (3.025, -1.75) {$3$};
		\node [style=wire label] (53) at (-2.725, 0.25) {$2$};
		\node [style=wire label] (54) at (-2.725, -1.75) {$2$};
		\node [style=gn] (55) at (6.65, -1.5) {};
		\node [style=gn] (56) at (6.65, 1.5) {};
		\node [style=none] (57) at (4.075, 1.5) {};
		\node [style=none] (58) at (4.075, 0) {};
		\node [style=none] (59) at (6.075, 0) {};
		\node [style=none] (60) at (6.075, 1.5) {};
		\node [style=wire label] (61) at (-0.95, -1.975) {$3$};
		\node [style=wire label] (62) at (-0.975, 0.275) {$3$};
		\node [style=none] (63) at (-3.525, -4) {$=$};
		\node [style=none] (64) at (-1.7, -4) {$\frac{\sqrt{3}}{2}$};
		\node [style=gn] (66) at (-2.275, 0) {};
		\node [style=none] (67) at (-7.2, -4) {$=$};
		\node [style=none] (68) at (-5.25, -4) {$\ldots$};
	\end{pgfonlayer}
	\begin{pgfonlayer}{edgelayer}
		\draw (2.center) to (3.center);
		\draw (4.center) to (1.center);
		\draw (5.center) to (1.center);
		\draw (7.center) to (8.center);
		\draw (11.center) to (6.center);
		\draw (12.center) to (6.center);
		\draw (13.center) to (7.center);
		\draw (17.center) to (14.center);
		\draw (15.center) to (16.center);
		\draw (19.center) to (18.center);
		\draw (6.center) to (7.center);
		\draw (21.center) to (22.center);
		\draw (23.center) to (20.center);
		\draw (24.center) to (20.center);
		\draw (26.center) to (27.center);
		\draw (30.center) to (25.center);
		\draw (31.center) to (25.center);
		\draw (32.center) to (26.center);
		\draw (25.center) to (26.center);
		\draw [in=60, out=180, looseness=0.75] (39.center) to (38);
		\draw [in=120, out=0, looseness=0.75] (42.center) to (41);
		\draw (38) to (37);
		\draw (37) to (41);
		\draw [bend right=90, looseness=1.50] (46) to (47);
		\draw [bend left=90, looseness=1.50] (50) to (51);
		\draw [in=180, out=-60, looseness=0.75] (38) to (40.center);
		\draw [in=0, out=-120, looseness=0.75] (41) to (43.center);
		\draw (42.center) to (47);
		\draw (43.center) to (36);
		\draw (51) to (40.center);
		\draw (48) to (56);
		\draw (56) to (60.center);
		\draw (59.center) to (50);
		\draw (58.center) to (39.center);
		\draw (57.center) to (46);
		\draw [in=180, out=0, looseness=1.25] (57.center) to (59.center);
		\draw [in=0, out=-180] (60.center) to (58.center);
	\end{pgfonlayer}
\end{tikzpicture}
}
\]
We represent the relevant basis states of binary trees of coupling spins and show how ZX rewrites allow us to derive expectation values of permutation operators directly from the structure of the network.

In condensed matter physics, we tackle symmetry-protected topological phases in 1D and 2D greatly simplifying previous diagrammatic computations on these systems Ref.~\cite{eastAKLTStatesZXDiagramsDiagrammatic2022}. For instance, the diagram
\[
\scalebox{0.75}{\input{figures/qufit-applications/2daklt-hexagon.tikz}}
\]
depicts the 2D AKLT Hexagon state.

We will show how the unitary spin-network gates of Ref.~\cite{eastSpinnetworksZXcalculus2022} can be compactly written in the form:
\[
\scalebox{0.75}{\input{figures/qufit-applications/su2-gate-example.tikz}}
\]
where the $\land$ box is shorthand explained in the text below. These diagrams are then integrated into previously derived diagrammatic QML calculations for use in gradient computations.

Finally, we will turn to Loop quantum gravity where we will diagrammatically calculate the minimal volume of a chunk of discretised space as $\frac{-\sqrt{3}}{4}$:

\[
	\scalebox{0.75}{\input{figures/qufit-applications/vol-state.tikz}}
\]
\[ =\quad
	\frac{-\sqrt{3}}{4}\scalebox{0.75}{\input{figures/qufit-applications/vol-min-intv3.tikz}}
\]

\subsection{Paper structure}

In concrete terms what is presented in this paper is as follows. We provide an introduction to how mixed-dimensional ZX calculus can be extended to allow us to focus on $SU(2)$ representations rather than qubits. We then explicitly introduce the Spin-ZX calculus with its generators and rewrite rules showing diagrammatically how the symmetrisation operator allows us to define different representations and connect them to qubits, detailing its rewrite properties and demonstrating diagrammatically that it has the expected behaviour. We then show how the diagrammatic description of spin-coupling via 3jm symbols or Clebsch-Gordan coefficients (these being isomorphic, with the first being invariant states and the second being the equivalent equivariant maps) can be formulated as mixed-dim ZX diagrams embedding them in the broader calculus. We then use the symmetriser-based rewrites to derive all the properties of these symbols diagrammatically. Given the embedding into mixed-dimensional ZX, we show how this allows us to formally express related mathematical objects in the algebra of $\mathfrak{su}(2)$ in a form that is related to but distinct from the Spin-ZX calculus itself.

To showcase the broad range of applications for these diagrams and to offer a starting point to other researchers, we then apply these techniques to a number of areas. Firstly, we start in the realm of quantum computing itself and derive permutational quantum computing (PQC) transition amplitudes entirely diagrammatically. This turns calculations usually requiring the use of more specialist Racah transformations, into the (diagrammatic) reduction of a tensor network. Next, we look to condensed matter physics, where we simplify and extend more cumbersome diagrammatic calculations for the analysis of 1D and 2D AKLT states previously attempted with qubit-only calculi~\cite{eastAKLTStatesZXDiagramsDiagrammatic2022}. Following this, we turn to calculations in quantum machine learning, demonstrating how parametrised quantum circuits designed for spin systems can be analysed, and providing as an example how to calculate the variance of the $SU(2)$ equivariant ansatz seen in Ref.~\cite{eastAllYou2023}. Finally, we apply our techniques to quantum gravity, demonstrating the ability to simplify the diagrammatic representation of the volume operator of loop quantum gravity, and computing the minimal quantised volume via an eigenvalue calculation performed entirely diagrammatically which simplifies and extends the results seen in Ref.~\cite{eastPhysiqueFormelleSpin2022}.

The general principle is that the Spin-ZX calculus streamlines the representation and manipulation of $SU(2)$ calculations, central to various domains including permutational quantum computing, loop quantum gravity, quantum machine learning, and condensed matter physics. This not only elucidates these calculations but gives a new perspective on the core relationships within $SU(2)$ representation theory and offers a formal diagrammatic language with which to analyse and apply them.

\section{ZX calculus for finite-dimensional quantum theory}

\label{sec:qufitzx}
In this section, we give an introduction to the general notion of a graphical calculus, and specifically, outline the parts of the mixed-dimensional ZX calculus~\cite{poorZXcalculusComplete2024} that we will need to outline the fragment that corresponds to the Spin-ZX calculus, which can be used to streamline the representation and manipulation of $SU(2)$ calculations.

\subsection{A general graphical calculus}\label{subsec:graphical-calculus}

A graphical calculus is a mathematical language for expressing and reasoning about processes using graphical representations~\footnote{Note here we are not using representation in the manner of representation theory, unfortunately, these two meanings do appear in this paper though context should indicate which is meant.} (i.e.\@ diagrams) whilst maintaining the rigour of an explicit interpretation.
By working with a graphical representation, we obtain higher expressivity, enabling a cleaner presentation of the underlying structure than if we were to rely solely on conventionally used symbolic notation.
This section explains the basic idea of a general graphical calculus; for more details, a thorough introduction to the topic is presented in~\cite{coeckeGeneralisedCompositionalTheories2016}.

A graphical calculus uses diagrams composed of boxes and wires, which can be manipulated according to a set of equalities called graphical rewrite rules.
The basic building blocks of a graphical language are its \emph{generators}:
\[
  \input{graphical-calculi/generators.tikz}
\]
With these generators, we can construct larger diagrams by \emph{composing} them sequentially or in parallel.
\emph{Sequential composition} (denoted by $\circ$) of two diagrams is given by connecting the outputs of one diagram to the inputs of the other:
\[
  \input{graphical-calculi/generalboxf.tikz}\ \circ \ \input{graphical-calculi/generalboxg.tikz}
  \quad \coloneqq \quad
  \input{graphical-calculi/sequelcomposition.tikz}
\]
\emph{Parallel composition} (denoted by $\otimes$) of two diagrams is given by placing two diagrams next to each other:
\[
  \input{graphical-calculi/generalboxf.tikz} \ \otimes \ \input{graphical-calculi/generalboxg.tikz}
  \quad \coloneqq \quad
  \input{graphical-calculi/parallelcomposition.tikz}
\]

In addition to composition, we can also reason about diagrams using graphical reasoning.
This is done by replacing sub-diagrams within a complex diagram according to a set of \emph{graphical rewrite rules}, that is, if
\[
  \input{graphical-calculi/rewritedm2LHS.tikz}
  \quad = \quad
  \input{graphical-calculi/rewritedm2RHS.tikz}
\]
then
\[
  \input{graphical-calculi/rewritedm1LHS.tikz}
  \quad = \quad
  \input{graphical-calculi/rewritedm1RHS.tikz}
\]
In particular, a lot of well-known graphical calculi, including the graphical calculus used in this paper, obey the following structural rules that allow us to move diagrams around freely:
\[
  \input{graphical-calculi/compactstructure-1.tikz}
\]
\[
  \input{graphical-calculi/compactstructure-2.tikz}
\]
\[
  \input{graphical-calculi/compactstructure-3.tikz}
\]

Lastly, to relate a graphical calculus to different mathematical systems, one needs to give an interpretation to diagrams such that sequential and parallel compositions are preserved.
In our case, the interpretation of diagrams is to C-linear maps, sequential composition is matrix multiplication, and parallel composition is the tensor product.

\subsection{Finite-dimensional ZX calculus}\label{subsec:qudit-zx}

We begin with an introduction to the generators of the finite-dimensional ZX calculus as it is presented in~\cite{poorCompletenessArbitraryFinite2023}, describing their interpretation as linear maps, and presenting useful notations.
Following this, we introduce the generators whose wires have mixed dimensions as given in~\cite{wangQufiniteZXcalculusUnified2022}, including the swap, and the mixed-dimensional Z-spider. The reason we turn to mixed-dimensional ZX at all instead of relying on variants of the standard ZX calculus is that representation theory involves the representation of the same group in different dimensional Hilbert spaces and studies how they interact. This naturally suggests that the most fitting framework to study this would be within a calculus-like mixed-dimensional ZX that is specifically tailored to this.

Note that all the ZX diagrams presented in this paper are read from top to bottom.
Furthermore, when working with mixed dimensions, each wire is labelled with its dimension, and only wires with the same dimension can be connected.
Since a wire labelled with dimension $1$ is just an empty diagram, throughout this paper each wire label will represent an integer strictly bigger than~$1$.

\subsubsection{Generators and their interpretation}\label{subsec:generators_n_interpretation}

We introduce the generators of the qudit ZX calculus and their standard interpretation as given in~\cite{poorZXcalculusComplete2024}.
We assume that each wire is labelled as a fixed positive integer $d$, thus for convenience, we ignore all the labels in this subsection.
The generators of the qudit ZXW calculus together with their standard interpretation $\interp{\cdot}$ are:
\begin{itemize}

  \item \emph{Z spider},
  \[
    \input{qufit-generators/mixed-zbox.tikz}
    \;\;\overset{\interp{\cdot}}{\longmapsto}\;\;
    \sum_{j=0}^{\min{\{d_i\}_i} - 1}\! a_j
    \ket{j,\cdots,j} \bra{j,\cdots,j}
    \]
    where $a_0 \coloneqq 1$ and $\overrightarrow{a} = (a_1, \cdots, a_{\min{\{d_i\}_i}-1})$.

    \item \emph{X spider},
    \[
      \input{qufit-generators/quditrspiderclassicnm.tikz}
    \;\; \overset{\interp{\cdot}}{\longmapsto}
    \sum_{
      \substack{
        i_1+\cdots+ i_m + j\\
        \equiv j_1+\cdots +j_n \! \Mod{d}
      }
    }\!
      \ket{i_1, \cdots, i_m}\bra{j_1, \cdots, j_n}
      \]
      where  $j \in \mathbb{Z}$.
      Two X spiders labelled $j$ and $j'$ are equal by definition if $j \equiv j' \! \Mod{d}$.
    If $0\leq j < d$, then X-spiders labelled $j$ that have a single input or output correspond to a computational basis costates/states as follows:
  \[
    \begin{tikzpicture}
	\begin{pgfonlayer}{nodelayer}
		\node [style={rn_phase}] (0) at (0, -0.5) {$K_j$};
		\node [style=none] (1) at (0, 0.75) {};
	\end{pgfonlayer}
	\begin{pgfonlayer}{edgelayer}
		\draw (0) to (1.center);
	\end{pgfonlayer}
\end{tikzpicture}

    \quad \overset{\interp{\cdot}}{\longmapsto} \quad
    \bra{j}
    \qquad \qquad
    \begin{tikzpicture}
	\begin{pgfonlayer}{nodelayer}
		\node [style={rn_phase}] (0) at (0, 0.75) {$K_j$};
		\node [style=none] (1) at (0, -0.5) {};
	\end{pgfonlayer}
	\begin{pgfonlayer}{edgelayer}
		\draw (0) to (1.center);
	\end{pgfonlayer}
\end{tikzpicture}

    \quad \overset{\interp{\cdot}}{\longmapsto} \quad
    \ket{d-j}
    \]

    \item The \emph{Hadamard box},
    \[
      \begin{tikzpicture}
	\begin{pgfonlayer}{nodelayer}
		\node [style=hbox] (0) at (0, 0) {$H$};
		\node [style=none] (1) at (0, -1) {};
		\node [style=none] (2) at (0, 1) {};
	\end{pgfonlayer}
	\begin{pgfonlayer}{edgelayer}
		\draw (2.center) to (0);
		\draw (1.center) to (0);
	\end{pgfonlayer}
\end{tikzpicture}

      \quad \overset{\interp{\cdot}}{\longmapsto} \quad
      \frac{1}{\sqrt{d}}\sum_{k, j=0}^{d-1}\omega^{jk}\ket{j}\bra{k}
    \]
    where $\omega = e^{i\frac{2\pi}{d}}$ is the $d$-th root of unity.
    Note that we could also use the other generators to define the Hadamard box (see~\cite[Rule (HD)]{poorZXcalculusComplete2024}); however, for the sake of simplicity, it is given as a generator here.

    \item The \emph{identity},
    \[
      \begin{tikzpicture}
	\begin{pgfonlayer}{nodelayer}
		\node [style=none] (1) at (1.0, 0.6) {};
		\node [style=none] (2) at (1.0, -0.6) {};
		\node [style=none] (3) at (1.0, -1.0) {};
		\node [style=none] (4) at (1.0, 1.0) {};
	\end{pgfonlayer}
	\begin{pgfonlayer}{edgelayer}
		\draw (1.center) to (2.center);
	\end{pgfonlayer}
\end{tikzpicture}
      \quad \overset{\interp{\cdot}}{\longmapsto} \quad
      I_d=\sum_{j=0}^{d-1}\ket{j}\bra{j}
      \]
      \item \emph{Swap},
      \[
        \input{qufit-generators/swap.tikz}
        \quad \overset{\interp{\cdot}}{\longmapsto} \quad
        \sum_{i=0}^{m-1}\sum_{j=0}^{n-1}\ket{j,i}\bra{i,j}
      \]
\end{itemize}

\subsubsection*{Notations}

For convenience, we introduce the following notation which will be used throughout the paper:
\begin{itemize}
  \item We omit the label of the spider when it corresponds to the all-ones vector.
  \begin{gather*}
    \input{definitions/circlegspiders-2.tikz}
    \quad \text{ where }
    \overrightarrow{1} = (1,\cdots,1).
  \end{gather*}
   \item We make two short notations with the Z spider whose  first/last $d-2$ components of the parameter vector are all zeros as follows:
  \[
    \input{definitions/lastditgbox.tikz}
  \]
  \[\input{definitions/firstditgbox.tikz}
  \]
  \[
    \text{ where }
    \quad
    a \in \mathbb C.
  \]


  \item Throughout the paper, we use the following notation:
  \[
    \overrightarrow{1_k} = \overbrace{(\underbrace{1,\dotsc, 1}_{k - 1},0, \dotsc, 0)}^{d}
    \qquad
    \text{ for some }
    d.
  \]

  \item 
  We define the cup and cap as follows:
  \begin{gather}
    \input{definitions/compactstructures-2.tikz}
    \qquad \qquad
    \input{definitions/compactstructures-1.tikz}
    \tag{S3}\label{rule:S3}\refstepcounter{equation}
  \end{gather}

  \item The phase-free X spider is defined as
  \[
    \input{definitions/pinkspiders-phasefree.tikz}
  \]


  \item We use a yellow $D$ box to denote the \emph{dualiser} as defined in~\cite{coeckeInteractingQuantumObservables2011}:
  \begin{gather}
    \input{definitions/dualiser.tikz}
    \qquad \overset{\interp{\cdot}}{\longmapsto} \qquad
    \sum_{i = 0}^{d \minu 1} \ket{d-i} \bra{i}
    \tag{Du}\label{rule:Du}\refstepcounter{equation}
  \end{gather}

  \item A multiplier~\cite{bonchiInteractingHopfAlgebras2017, caretteSZXCalculusScalableGraphical2019, boothCompleteZXcalculi2022} labelled by $m$ indicates the number of connections between green and pink nodes.
  \begin{gather}
    \input{definitions/multiplier.tikz}
    \qquad \qquad
    \input{qufit-generators/multiplier-mod.tikz}\\
    \tag{Mu}\label{rule:Mu}\refstepcounter{equation}
    \input{definitions/multiplier-t.tikz}
  \end{gather}

 \end{itemize}

\subsection{ZX rules prerequisite for Spin-ZX calculus}

ZX ultimately uses the same qubit dimension for all its wires. Though practical for qubit engineering purposes for broader theoretical physics it is often the case that different dimensional quantum systems interact. For this reason, a calculus designed specifically for the representation of spin must be able to succinctly describe the interaction of any finite-dimensional quantum systems. We present a set of ZX rules for the ZX calculus for finite-dimensional quantum theory here:
\begin{gather}
  \input{qufit-axioms/qufit-gengspiderfusedit-2col.tikz}
  \tag{S1}\label{rule:S1}\refstepcounter{equation}
\end{gather}
\vspace{-0.5cm}
\begin{flalign*}
  \text{where}\quad
  m = \min_{t = 1}^j m_t, \quad
  n = \min_{t = 1}^\ell n_t, \quad
  r = \min_{t = 1}^s r_t,\\
  \protect\overrightarrow{a}=(a_1, \dotsc, a_{m-1}),
  \protect\overrightarrow{b}=(b_1, \dotsc, b_{n-1}),\\
  M = \min\{m, n, r\},\quad \text{and} \quad
  \protect\overrightarrow{ab'}=(a_1 b_1,..., a_{M-1} b_{M-1}, 0,..., 0).
\end{flalign*}
\begin{gather}
  \input{axioms/redspider0pfusedit2-2col.tikz}
   \tag{S4}\label{rule:S4}\refstepcounter{equation}
\end{gather}
  \noindent
  \allowdisplaybreaks
  \begin{gather}
    \input{axioms/s2qudit.tikz}
    \tag{S2}\label{rule:S2}\refstepcounter{equation} \\
    \input{axioms/b2qudit.tikz}
    \tag{B2}\label{rule:B2}\refstepcounter{equation}\\
    \input{axioms/rdotaemptydit0.tikz}
    \tag{Ept}\label{rule:Ept}\refstepcounter{equation} \\
    \input{axioms/color.tikz}
    \tag{HZ}\label{rule:HZ}\refstepcounter{equation} \\
    \begin{tikzpicture}
	\begin{pgfonlayer}{nodelayer}
		\node [style=gbox] (0) at (-1, 0.5) {$0$};
		\node [style=rn] (1) at (1, 0.5) {};
		\node [style=none] (2) at (-1, -0.5) {};
		\node [style=none] (3) at (1, -0.5) {};
		\node [style=none] (4) at (0, 0) {=};
	\end{pgfonlayer}
	\begin{pgfonlayer}{edgelayer}
		\draw (2.center) to (0);
		\draw (1) to (3.center);
	\end{pgfonlayer}
\end{tikzpicture}

    \tag{Zer}\label{rule:Zer}\refstepcounter{equation} \\
    \input{qufit-axioms/qufit-k1copy.tikz}
    \tag{K0}\label{rule:K0}\refstepcounter{equation}
  \end{gather}
  where $ 0 \leq j \leq m-1, N = \min\{m, n_1, n_2\}$
  \begin{gather}
    \input{axioms/k2adit.tikz}
    \tag{K2}\label{rule:K2}\refstepcounter{equation}
  \end{gather}
  where $\displaystyle \protect{k_j(\overrightarrow{a})}=\left(\frac{a_{1-j}}{a_{d-j}}, \dotsc, \frac{a_{d-1-j}}{a_{d-j}}\right).$
  \begin{gather}
    \input{axioms/p1sdit2.tikz}
    \tag{P1}\label{rule:P1}\refstepcounter{equation} \\
    \input{axioms/dcomwtha0.tikz}
    \tag{D1}\label{rule:D1}\refstepcounter{equation}
  \end{gather}
  where $\protect\overleftarrow{a}=(a_{d-1}, \dotsc, a_1)$
  \begin{gather}
  \input{axioms/zphasemerge.tikz}
  \tag{PA}\label{rule:pa}\refstepcounter{equation}
\end{gather}
where $r_0=\sum_{\substack{i = 0}}^{d-1}  a_i b_{d \minu i \! \Mod{d}}$,  the $k^{\text{th}}$ element of $\overrightarrow{c}$ is $\displaystyle c_k~=~\frac{1}{r_0}\sum_{\substack{i = 0}}^{d-1}  a_i b_{k \minu i \! \Mod{d}}$
  \begin{gather}
    \input{qufit-axioms/dimInc.tikz}
    \tag{DA}\label{rule:DA}\refstepcounter{equation}\\
    \input{qufit-axioms/phasecopymixd.tikz}
    \tag{PC}\label{rule:PC}\refstepcounter{equation}
  \end{gather}
   if $p_i p_j = q_{i + j \! \Mod{d}}$ for all $1 \leq i < m$ and $1 \leq j < n$.
  \begin{gather}
    \input{qufit-axioms/x-dim-change.tikz}
    \tag{DZX}\label{rule:DZX}\refstepcounter{equation}\\
      \input{qufit-axioms/xabsorbed.tikz}
    \tag{XB}\label{rule:XB}\refstepcounter{equation}
  \end{gather}
    where $m > n$.
  \begin{gather}
    \input{qufit-axioms/kdcommutev2.tikz}
   \tag{PK}\label{rule:PK}\refstepcounter{equation} 
  \end{gather}
  where $2\leq n \leq m$.


\section{Spin-ZX calculus}\label{sec:penrose}
\allowdisplaybreaks
\setlength{\jot}{5pt}

In this section, we introduce the Spin-ZX calculus itself, which we elevate to the level of a formal diagrammatic language by embedding it into to the ZX calculus as the fragment that describes the recoupling theory of $SU(2)$.
This fragment encompasses Yutsis diagrams, and their compositions, which are often called \emph{spin networks}~\cite{majorSpinNetwork1999}.
A brief introduction to Yutsis diagrams can be found in Appendix~\ref{sec:yutsis}.

The generators of the Spin-ZX calculus are given in \cref{tab:penrose-generators}, also given in the left half of \cref{fig:generators-intro}.
\begin{table*}[ht]
  \centering
  \begin{tabular}{c|c|c}
  \hline
  Yutsis diagram & ZX diagram & Interpretation \\
  \hline
  $\scalebox{.8}{\input{figures/yutsis/identity.tikz}}$ & $\scalebox{.8}{\input{figures/yutsis/identity-zx.tikz}}$ & {\footnotesize $\displaystyle \sum_{m} \ket{j;m}\bra{j;m}$} \\
  \hline
  $\scalebox{.8}{\input{figures/yutsis/3j-state.tikz}}$ & $\scalebox{.8}{\input{figures/yutsis/3j-state-zx.tikz}}$ & {\footnotesize $\displaystyle \sum_{m_1,m_2,m_3} \mqty(j_1 & j_2 & j_3 \\ m_1 & m_2 & m_3) \ket{j_1;m_1,\, j_2;m_2,\, j_3;m_3}$} \\
  \hline
  $\scalebox{.8}{\input{figures/yutsis/cup-ii-spin-j.tikz}}$ & $\scalebox{0.8}{\input{figures/yutsis/cup-ii-spin-j-zx.tikz}}$ & {\footnotesize $\displaystyle \sqrt{2j \plus 1}\sum_{m} (-1)^{j+m+1} \bra{j;m, \, j;-m}$} \\
  \hline
  \end{tabular}
  \caption{Generators of the Spin-ZX calculus. Here, $j, j_1, j_2, j_3 \in \mathbb{N}/2$, $x_{lm} = j_l + j_m - j_n$, and $\protect\overrightarrow{1^{-}} = \left( (-1)^1, \cdots, (-1)^{2j} \right)$. $ j_1, j_2, j_3 $ satisfy the Clebsch-Gordan conditions~\eqref{eq:CGcondition}.  The normalisation factor $N(j_1,j_2,j_3)$ is given by \cref{eq:N-factor}.}
  \label{tab:penrose-generators}
\end{table*}
Using these generators, we can define other orientations of the 3-valent vertex as shown below:
\begin{equation*}
  \scalebox{.75}{\input{figures/yutsis/3-valent-notations-1.tikz}}
\end{equation*}
\begin{equation*}
  \scalebox{.75}{\input{figures/yutsis/3-valent-notations-2.tikz}}
\end{equation*}
\begin{equation*}
  \scalebox{.75}{\input{figures/yutsis/3-valent-notations-3.tikz}}
\end{equation*}
We can define the vertex that embeds spin $\frac{n}{2}$ into $n$ copies of the fundamental irrep by stacking the 3-valent vertices as shown below:
\begin{equation*}
  \scalebox{.75}{\input{figures/yutsis/embedding-1.tikz}}
\end{equation*}
\begin{equation*}
  \scalebox{.75}{\input{figures/yutsis/embedding-2.tikz}}
\end{equation*}
Composing the embedding vertex with its adjoint, we can define the symmetriser:
\begin{equation*}
  \scalebox{.75}{\input{figures/yutsis/symmetriser.tikz}}
\end{equation*}

If $j_1 = j_3 = \frac{1}{2}, j_2 = 0$, then we have 
\begin{equation*}
\input{figures/yutsis/bellstateyut.tikz} \longmapsto \input{figures/yutsis/bellstatezx.tikz}
\end{equation*}
or in the conjugate case
\begin{equation*}
	\input{figures/yutsis/belleffyut.tikz} \longmapsto \input{figures/yutsis/belleffzx.tikz}
\end{equation*}

Next, we provide the rules of the Spin-ZX calculus, which can be derived from the rules of the ZX calculus.
\begin{mdframed}\label{Table:Spin-ZX-rewrites}
  \begin{gather}
    \scalebox{0.9}{\input{figures/yutsis/even-permutation-rule-2col.tikz}} \tag{even-permute} \label{rule:even-permutation} \refstepcounter{equation} \\
    \scalebox{0.9}{\input{figures/yutsis/odd-permutation-rule-2col.tikz}} \tag{odd-permute} \label{rule:odd-permutation} \refstepcounter{equation} \\
    \scalebox{0.9}{\input{figures/yutsis/3j-to-penrose-2col.tikz}} \tag{3j-decomposition} \label{rule:3j-decomposition} \refstepcounter{equation} \\
    \scalebox{0.9}{\input{figures/yutsis/isometry.tikz}} \tag{isometry} \label{rule:isometry} \refstepcounter{equation} \\
    \scalebox{0.9}{\input{figures/binor/projector.tikz}} \tag{projector} \label{rule:projector} \refstepcounter{equation} \\
    \scalebox{0.9}{\input{figures/binor/stacking.tikz}} \tag{stacking} \label{rule:stacking} \refstepcounter{equation} \\
    \scalebox{0.9}{\input{figures/binor/cappingv3.tikz}} \tag{capping} \label{rule:capping} \refstepcounter{equation} \\
    \scalebox{0.9}{\input{figures/binor/slidingv2.tikz}} \tag{sliding} \label{rule:sliding} \refstepcounter{equation} \\
    \scalebox{0.9}{\input{figures/binor/loopingv2.tikz}} \tag{looping} \label{rule:looping} \refstepcounter{equation}
  \end{gather}
\end{mdframed}
The rest of the section will motivate the generators and derive the rules of the Spin-ZX calculus.


\subsection{\texorpdfstring{$SU(2)$}{SU(2)} Representation theory}
\label{sec:SU(2)_representation}

Let us now understand the representation theory of SU(2) better in order to understand the fragment more clearly and along the way outline how it embeds into ZX. The Lie group SU(2) has importance in many areas in physics.
It is most well-known for the theory of quantum angular momentum, but it is also central to the study of isospins, electroweak theory, and qubit quantum computation.
$SU(2)$ is the group of $2\times 2$ unitary matrices with determinant $1$.
A \emph{unitary representation} of $SU(2)$ is a group homomorphism $SU(2)\rightarrow U(\mathcal{H})$ where $U(\mathcal{H})$ denotes the set of unitaries on a (finite-dimensional) Hilbert space $\mathcal{H}$.
We can decompose these as a direct sum of \emph{irreducible representations} (irreps).
Each such representation corresponds to a dimension, and for every dimension there is an irrep.
We label these irreps by half-integers $j \in \mathbb{N}/2$ which in physics are better known as \emph{spins}.
The spin-$j$ irrep has dimension $2j+1$ and is unique up to isomorphism.

\paragraph{The irreps of SU(2).}
The \emph{fundamental} irrep (the spin-$1/2$ representation) $SU(2) \to U(\mathbb{C}^2)$ is simply the $2\cross2$ matrix multiplication of $SU(2)$ itself over $\mathbb{C}^2$.
We can then define the higher-spin irreps by using `symmetrised copies' of this representation. 
First, define the \emph{symmetrisation projection} where for $n=2j$ we write~\footnote{Here we encounter the mathematicians labelling of the irreps by integers versus the physicists half integers, the difference is a result of the actual measurement outcomes observed in spin experiments which includes a factor of $\frac{1}{2}$.} $\mathcal{S}_{n}$ as the endomorphism over $(\mathbb{C}^2)^{\otimes n}$ such that
\begin{equation}
	\label{eq:symmetrisation_projector}
	\mathcal{S}_{n}(v_1 \otimes \cdots \otimes v_{n}) = \frac{1}{(n)!} \sum_{\sigma \in \mathfrak{S}_{n}} U_\sigma \left(  v_1 \otimes ... \otimes v_{n} \right).
\end{equation}
Here $\mathfrak{S}_{n}$ is the $2j$-element permutation group and the $U_\sigma$ is the \emph{permutation unitary}~\footnote{Which permutes the labels $U_\sigma \left(  v_1 \otimes \cdots \otimes v_{n} \right) \overset{\text{def}}=   v_{\sigma(1)} \otimes \cdots \otimes v_{\sigma(n)}$.}.
The action of $\mathcal{S}$ is to send every vector to a symmetric vector that is invariant under the action of the permutation unitaries.
It is straightforward to check that $\mathcal{S}$ is indeed a projector (i.e.~self-adjoint and idempotent).
It then defines a subspace of $(\mathbb{C}^2)^{\otimes n}$ that we denote by $\mathcal{H}_j$.
This Hilbert-space consists of the symmetric vectors and has dimension $2j+1$ (i.e.\@ $n+1$).
Since $\mathcal{H}_j$ is a subspace of $(\mathbb{C}^2)^{\otimes n}$, there exists an isometry $V_j: \mathcal{H}_j \to (\mathbb{C}^2)^{\otimes n}$ which embeds the canonical basis of $\mathcal{H}_j$ into $(\mathbb{C}^2)^{\otimes n}$.
\begin{restatable}{proposition}{isometrydef}\label{prop:isometry-def}
    This isometry is given by the following diagram:
    \begin{equation}
        \input{figures/binor/isometry-def.tikz}\label{eq:isometry}
    \end{equation}
    where $\overrightarrow{\sqrt{a_n}} = \left( \frac{1}{\sqrt{\binom{n}{1}}}, \cdots, \frac{1}{\sqrt{\binom{n}{k}}}, \cdots, \frac{1}{\sqrt{\binom{n}{n}}} \right).$
\end{restatable}
We can verify that this diagram is indeed an isometry:
\begin{proposition}
    \label{prop:isometry}
    $V_j^\dagger V_j = I$.
    That is,
    \begin{equation}
        \input{figures/binor/isometry.tikz}
    \end{equation}
\end{proposition}
\begin{proof}
    It follows from~\eqref{rule:S1} and \cref{merge2}.
\end{proof}
With the above defined, we can represent the symmetriser operator $\mathcal{S}_n$ diagrammatically with the following ZX diagram. This composition of the isometry and its adjoint $V_j V_j^\dagger = \mathcal{S}_n$ is precisely the projection onto the subspace $\mathcal{H}_j$.
\begin{restatable}{proposition}{symmetriser}
    \begin{equation}
        \label{symmetriser}
        \input{figures/binor/sym-j.tikz}
        \quad \longmapsto
        \input{figures/qufit-applications/spinjiireps2.tikz}
    \end{equation}
    where
    \begin{equation*}
        n = 2j,
        \ \overrightarrow{a_n} = \left( \frac{1}{\binom{n}{1}}, \cdots, \frac{1}{\binom{n}{k}}, \cdots, \frac{1}{\binom{n}{n}} \right).
    \end{equation*}
\end{restatable}
\begin{proof}
    It follows from \cref{prop:isometry-def} and spider fusion \eqref{rule:S1}.
\end{proof}

We can write the canonical orthonormal basis of $\mathcal{H}_j$ as
\begin{equation}\label{eq:canon-basis}
	\ket{j;m} \ \coloneqq \  \sqrt{\binom{2j}{j-m}} \, \mathcal{S}_{n} \underbrace{\left( \ket{0} \otimes ... \otimes \ket{1} \right)}_{\text{$j-m$ times 1}}
\end{equation}
with $m \in \{ -j,-j+1,\ldots, j-1, j \}$.
Note that here we are using the physicists notation with $\ket{j;m}$ denoting intrinsic angular momentum $j$ and its azimuthal component $m$ of a single state, this is not a tensor product.
In addition, we also have spin-$0$ state which functions as the cup/cap mapping serving as the conjugate of the fundamental irrep spin-$\frac{1}{2}$ space:
\begin{equation}
    \ket{0;0}
    \ \ =\ \ \frac{1}{\sqrt{2}}\ \input{figures/binor/cupcap.tikz}
    \ \ =\ \ \frac{1}{\sqrt{2}}(\ket{01} - \ket{10})
\end{equation}
In reality this state is a trivial form of the 3JM symbol described below in \cref{3jm-sec} where one of the $j=0$, which is why is has different (anti)symmetry properties, we add it here for completeness from the physicists' perspective.

Consider the spin-$\frac{3}{2}$ basis, which is
$\left\{\ket{\frac{3}{2};\frac{3}{2}},\ket{\frac{3}{2};\frac{1}{2}},\ket{\frac{3}{2};-\frac{1}{2}},\ket{\frac{3}{2};-\frac{3}{2}}\right\}$, as a more general example of the canonical basis.
Written in terms of a qubit basis, this is
\begin{align*}
    \left\{\ket{000},\frac{1}{\sqrt{3}} \left( \ket{001}+\ket{010}+\ket{100} \right)\right.,\\
    \left.\frac{1}{\sqrt{3}} \left( \ket{011}+\ket{101}+\ket{110} \right),\ket{111}\right\}
\end{align*}
These correspond to the following diagrams~\footnote{We have omitted scalars for brevity.}:
\[
    \input{figures/binor/ket3232.tikz}
    \qquad\quad
    \input{figures/binor/ket3212.tikz}
\]
\[
    \input{figures/binor/ket32-12.tikz}
    \qquad\quad
    \input{figures/binor/ket32-32.tikz}
\]
Alternatively, using~\eqref{eq:isometry} and~\eqref{eq:canon-basis}, we can obtain the following diagrammatic representation of the canonical basis $\ket{j;m}$ of $\mathcal{H}_j$:
\begin{equation}
    \input{figures/binor/canonical-basis-2col.tikz}
\end{equation}

We discuss some properties of the symmetriser $\mathcal{S}_n$ as given in~\cite[Section 3.6]{petersonTraceDiagrams2006} using the Spin-ZX calculus.
They are shown in \cref{fig:symmetriser-properties} and proved in \cref{sec:symmetriser-proofs}.

\begin{center}
    \begin{mdframed}
            \begin{restatable}[Projector]{proposition}{symmetriseridep}
                \label{prop:symmetriseridep}
            $\mathcal{S}_n^2=\mathcal{S}_n$, that is,
            \[
                \input{figures/binor/projector.tikz}
            \]
            \end{restatable}
            \begin{restatable}[Invariance]{proposition}{invariance}
                \label{prop:invariance}
            For an arbitrary $2 \times 2$ unitary represented by \input{figures/qufit-applications/Ubox.tikz}, the following equality holds:
            \[
                \input{figures/binor/matrixcommute.tikz}
            \]
            \end{restatable}
            \begin{restatable}[Stacking]{proposition}{stacking}
            Suppose $k \leq j$. Then
            \[
                \input{figures/binor/stacking.tikz}
            \]
            \end{restatable}
            \begin{restatable}[Capping / Cupping]{proposition}{cappingcupping}
            \[
                \input{figures/binor/cappingv2.tikz}
            \]
            \end{restatable}
            \begin{restatable}[Symmetriser Sliding]{proposition}{symmetrisersliding}
            \[
                \input{figures/binor/slidingv2.tikz}
            \]
            \end{restatable}
            \begin{restatable}[Looping Relation]{proposition}{loopingrelation}
                \label{prop:loopingrelation}
            \[
                \input{figures/binor/loopingv2.tikz}
            \]
            \end{restatable}
    \end{mdframed}
    
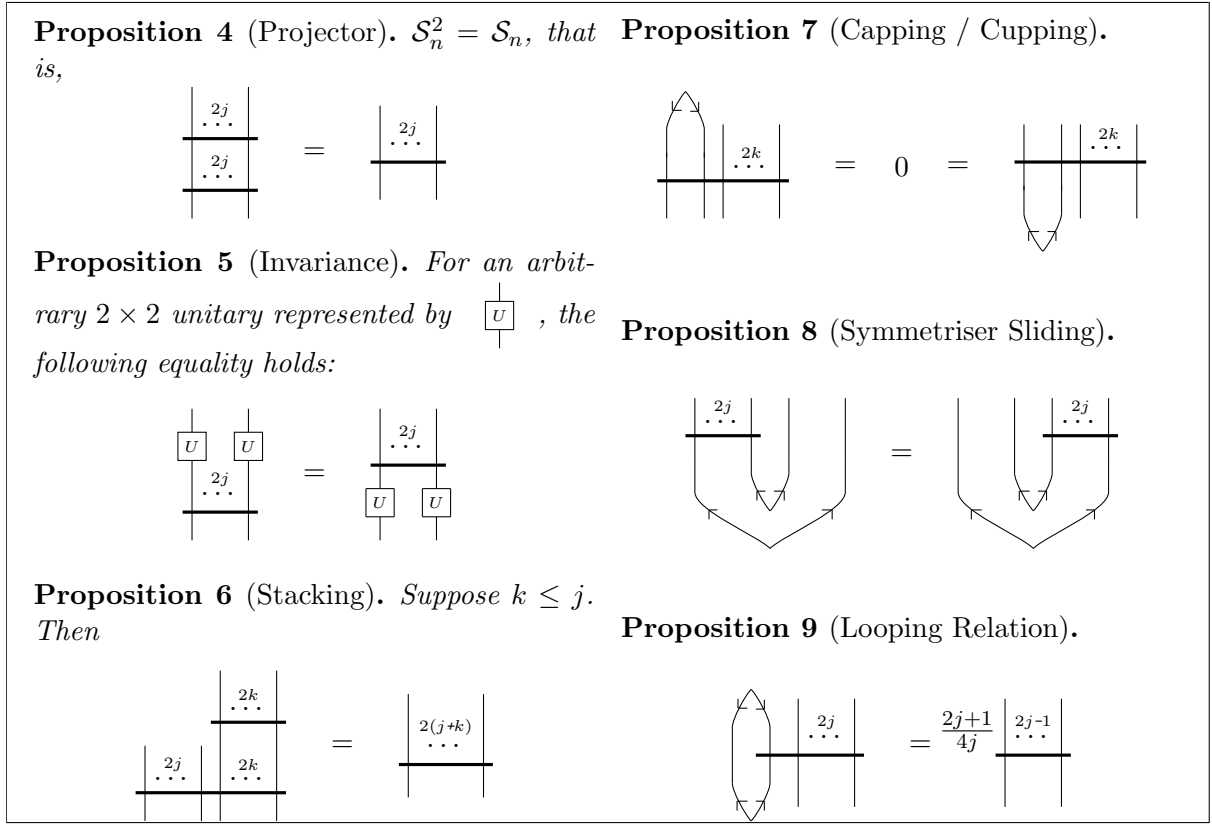
\captionof{figure}{Symmetriser properties}
    \label{fig:symmetriser-properties}
\end{center}

\paragraph{Diagrammatic representation of the Wigner Matrix}

With the isometry (\ref{eq:isometry}) and its dagger, we can represent the Wigner matrix $D^j_u$ as follows:
\begin{equation}\label{wignermat}
    \input{figures/binor/wignermatrix.tikz}
\end{equation}
where $u \in SU(2)$.

To prove the correctness of the diagram (\ref{wignermat}), we just need to show the following:

\begin{restatable}{proposition}{wignermatrixelement}
\[
D^j_{mn}(u) = \input{figures/binor/wignermatrixelem.tikz} = E,
\]
\[
    \text{where } u= \left(\!
    \begin{array}{ccc}
        a & b \\
        c & d
    \end{array}
    \!\right)\!, F = a^{j+n -k}b^{m-n+k}c^kd^{j-m -k}\!,\text{ and}
\]
\[
    E =
    \sum_{k} \frac{ \sqrt{(j-m)!(j+m)!(j-n)!(j+n)!} }{k!(j- m- k)!(j+n -k)!(m-n+k)!}F
\]
where the sum runs over all the values of $k$ for which the argument of every factorial is
non-negative~\cite{makinenIntroductionSU22019}.

\end{restatable}

\subsection{Spin coupling}\label{sec:3jsymbols}

Due to linearity, we can define an action of a group element $U \in SU(2)$ on $\mathcal{H}_j$ as follows:
\begin{equation}\label{eq:group-action}
	U \cdot (v_1 \otimes \dots \otimes v_{2j}) = (U v_1) \otimes \dots \otimes (U v_{2j}).
\end{equation}
where $Uv_j$ is the usual matrix multiplication of the $2\times 2$ matrix $U$ with the $\mathbb{C}^2$ vector $v_j$.
The corresponding action of a Lie algebra element, $a \in \mathfrak{su}(2)$, is then given as
\begin{equation}
	a \cdot (v_1 \otimes \dots \otimes v_{2j}) = \sum_{k=1}^{2j} v_1 \otimes \dots \otimes (a v_k ) \otimes \dots \otimes v_{2j}.
\end{equation}

These  are the fundamental building blocks from which other representations are built.
Any finite-dimensional representation of $SU(2)$ is completely reducible, meaning it can be written as a direct sum of irreps.
Specifically, a tensor product of irreps can be decomposed into a direct sum of irreps via a bijective intertwiner (or equivariant map).
Recoupling theory aims to study the space of such intertwiners.

\paragraph{Clebsch-Gordan coefficients.}
Given $\mathcal{H}_{j_1}$ and $\mathcal{H}_{j_2}$, their tensor product $\mathcal{H}_{j_1} \otimes \mathcal{H}_{j_2}$ can be decomposed into a direct sum of irreps with the following equivalence of representations
\begin{equation}
	C : \mathcal{H}_{j_1} \otimes \mathcal{H}_{j_2} \longrightarrow \bigoplus_{j=|j_1-j_2|}^{j_1+j_2} \mathcal{H}_j,
\end{equation}
Specifically for any state $\ket{j;m} \in \mathcal{H}_{j}$ we can write
\begin{align}\label{injectionmap}
	|j;m\rangle
  &=\sum_{m_1=\minu j_1}^{j_1}\sum_{m_2=\minu j_2}^{j_2}\! |j_1;m_1.j_2;m_2\rangle\langle j_1;m_1.j_2;m_2|j;m\rangle\\
  &=\sum_{m_1=\minu j_1}^{j_1}\sum_{m_2=\minu j_2}^{j_2}\! C^{jm}_{j_1m_1j_2m_2} |j_1;m_1.j_2;m_2\rangle
\end{align}
where $C^{jm}_{j_1m_1j_2m_2}$ are the \textit{Clebsch-Gordan coefficients} defined as:
\begin{equation}
	C^{jm}_{j_1m_1j_2m_2}  \overset{\text{def}}= \braket{j_1;m_1 . j_2;m_2}{j;m}.\label{EQ:CG}
\end{equation}
The indices can be read as tensor indices, with $m \in \{-j,\dots,j\}$ labelling the elements of a canonical basis of $\mathcal{H}_j$ (and similarly for $m_1$ and $m_2$). For all the coefficients we have $C^{jm}_{j_1m_1j_2m_2} \in \mathbb{R}$.
It turns out that if $m \neq m_1+m_2$, then also $C^{jm}_{j_1m_1j_2m_2} =0$.
Additionally, by convention we set $C^{jm}_{j_1m_1j_2m_2} = 0$ when the \textit{Clebsch-Gordan conditions} are not satisfied:
\begin{equation}\label{eq:CGcondition}
	\begin{split}
		j_1+j_2+j \in \mathbb{N} \\
		|j_1-j_2|\leq j \leq j_1+j_2.
	\end{split}
\end{equation}

\subsection{Embedding the generators of the Spin-ZX calculus}\label{3jm-sec}

In this section, we show how the primary generator of the Spin-ZX calculus, which is the 3-valent vertex of Yutsis diagrams is presented as a mixed-dimensional ZX diagram.
This vertex, when all edges are pointing outwards, represents a state called the 3j-state.
This state lives in the tensor product of three Hilbert spaces and it is invariant under the action of $SU(2)$.

The description of the Clebsch-Gordan coefficients has an asymmetry between the Hilbert spaces being tensored, and the Hilbert spaces it is being decomposed into. If we describe the situation in terms of a single state composed of three Hilbert spaces rather than privileging two spaces as inputs of a map to the third the situation becomes more symmetric.

Suppose the Clebsch-Gordan conditions~\eqref{eq:CGcondition} are satisfied (for $j=j_3$), then
\begin{equation}
	\dim \text{Inv}_{SU(2)} ( \mathcal{H}_{j_1} \otimes \mathcal{H}_{j_2} \otimes \mathcal{H}_{j_3}) = 1.
\end{equation}
where we understand $\text{Inv}_{SU(2)}(\mathcal{H})$ as a subspace of the given Hilbert space $\mathcal{H}$ that is invariant under the action of a representation of $SU(2)$ (more exactly we should state which representation but this clear from context). Hence, there is a unique unit vector in this subspace, up to phase.

To obtain this vector, we first observe that the following cup diagram is invariant under the action of $SU(2)$, i,e.\@ for any $u \in SU(2)$:
\begin{align}
  \input{figures/binor/g-on-cupszx-l.tikz}
  \quad&\cong\quad
  u \cdot (\ket{01} - \ket{10}) \nonumber \\
  = (\ket{01} - \ket{10})
  \quad&=\quad
  \input{figures/binor/g-on-cupszx-r.tikz}
\end{align}
Then the diagram
\begin{align}\label{3jstatedef}
	&\ket{j_1,j_2,j_3} \quad \overset{\text{def}}= \nonumber \\
  &\frac{(-1)^{j_1-j_2+j_3}}{N(j_1,j_2,j_3)} \scalebox{.8}{\input{figures/binor/3-valent-binorzx.tikz}}
\end{align}
depicts a vector that belongs to the one dimensional space $\text{Inv}_{SU(2)} ( \mathcal{H}_{j_1} \otimes \mathcal{H}_{j_2} \otimes \mathcal{H}_{j_3})$, which can be justified by the fact that the vector remains unchanged after plugging in the diagrams of Wigner matrices  (\ref{wignermat}) with the same $u \in SU(2)$ as shown in \cref{prop:statethenwignermatrices}.
Here the normalisation factor $N(j_1,j_2,j_3)$ is given by \cref{eq:N-factor}.
This should be interpreted as a kind of \enquote{railroad switch}, where the fundamental wires within the three symmetrised bundles redistribute between themselves.
Because we are dealing with symmetrised spaces, we only care about the number of wires going from each bundle to the other bundle.
It turns out that there is only one way to connect the wires when it is not impossible.
The Clebsch-Gordan conditions~\eqref{eq:CGcondition} precisely state when such a recoupling is possible.

\begin{restatable}{proposition}{statethenwignermatrices}
  \label{prop:statethenwignermatrices}
  \[
    \scalebox{.8}{\input{figures/binor/g-on-cupszx-r-then-wigner-matrix-lem-2col.tikz}}
  \]
\end{restatable}

Now we derive a simplified ZX diagram which represents the state $\ket{j_1,j_2,j_3}$.
This 3j state corresponds to the basic generator of Yutsis diagrams, which is a 3-valent vertex.
\begin{restatable}{theorem}{threejstate} \label{thm:3jstate}
    \begin{equation}\label{3jstatedgm}
        \scalebox{0.75}{\input{figures/3jsymbolcalc/3jmap-alt-2v3-2col.tikz}}
    \end{equation}
  where $x_{12}=j_1+j_2-j_3,\  x_{13}=j_1+j_3-j_2,\  x_{23}=j_2+j_3-j_1, \text{ and } \overrightarrow{\frac{1}{a^{-}}}
      =
      \left(
        (-1)^1\binom{x_{k \ell}}{1}, \cdots, (-1)^{x_{k \ell}}\binom{x_{k \ell}}{x_{k \ell}}
      \right).
$
\end{restatable}
The arrows pointing out of the vertex indicate it is an output.
A general 3-valent vertex can also have inputs, which are represented by wires entering the vertex.
Changing the direction of the wires corresponds to negating the magnetic basis, along with some phase.
We can change the wire directions by composing with the ZX diagram shown below, which is an analogue of the singlet effect.
\begin{equation}
  \scalebox{0.75}{\input{figures/yutsis/inversionvertex.tikz}}
\end{equation}
where $\overrightarrow{1^{-}} = \left( (-1)^1, \cdots, (-1)^{2j} \right).$
Composing this diagram with the 3j-state gives us the injection map given in \cref{injectionmap}.
\begin{restatable}{proposition}{injectionmap}
  The injection $\mathcal{H}_{j_3} \longrightarrow \mathcal{H}_{j_1} \otimes \mathcal{H}_{j_2} $ as given in \cref{injectionmap} can be represented by the following diagram:
    \begin{equation}\label{injectiondgm}
  \scalebox{0.75}{\input{figures/3jsymbolcalc/cgcoefficientv2-2col.tikz}}
   \end{equation}
\end{restatable}
In general, the composition of 3j-states naturally gives a network of interacting spin spaces, which is often termed a \emph{spin network}.
For example, the four-valent intertwiner can be obtained by composing two 3-valent vertices as follows:
\[
  \scalebox{.75}{\input{figures/3jsymbolcalc/3jmcomposition-2col.tikz}}
\]
In the rest of the paper, we will omit the arrows on the edges and use the orientation of the diagrams to indicate input and output wires.

\subsection{Wigner's \texorpdfstring{$3jm$}{3jm}-symbol.}
We can decompose the 3j vector in the magnetic basis as:
\begin{align}\label{eq:definition jstate}
	&\ket{j_1,j_2,j_3} \ = \nonumber \\
  &\sum_{m_1,m_2,m_3} \mqty(j_1 & j_2 & j_3 \\ m_1 & m_2 & m_3) \ket{j_1;m_1,\, j_2;m_2,\, j_3;m_3},
\end{align}
where the coefficients are called the \textit{Wigner's $3jm$-symbol}.
We can derive their explicit form by plugging in $\ket{j_1;m_1,\, j_2;m_2,\, j_3;m_3}$ in \cref{thm:3jstate}.
\begin{restatable}{theorem}{threejsymbol} \label{thm:3jsymbol}
  Suppose $m_1+m_2+m_3=0$, and $j_1, j_2, j_3$ satisfy the Clebsch-Gordan conditions.
  Then, the $3jm$-symbol is given by the following
  \[
    \left(
      \begin{array}{ccc}
        j_1 & j_2 & j_3 \\
        m_1 & m_2 & m_3
      \end{array}
    \right)
    \;\;=\;\;
    \scalebox{0.75}{\input{figures/3jsymbolcalc/3jm-symbolv2.tikz}}
  \]
  The diagram above evalutes to the following expression for the $3jm$-symbol from~\cite{shorePrinciplesAtomic1967}:
    \begin{align*}
     &(-1)^{j_1-j_2-m_3}\sqrt{\frac{(j_2+j_3-j_1)!(j_1+j_3-j_2)!(j_1+j_2-j_3)!}{(j_1+j_2+j_3+1)!}} \vspace{0.3cm}\\
    &\times \sqrt{(j_1-m_1)!(j_1+m_1)!(j_2-m_2)!} \vspace{0.3cm}\\
    &\times \sqrt{(j_2+m_2)!(j_3-m_3)!(j_3+m_3)! } \vspace{0.3cm} \\
    &\times  \sum_{k=K}^N \left(\frac{(-1)^{k}}{k!(j_1+j_2-j_3-k)!(j_1- m_1 - k)!} \vspace{0.3cm} \right.\\
    &\times \left. \frac{1}{(j_3-j_2+m_1 +k)!(j_3-j_1-m_2+k)!(j_2+m_2-k)!}\right)
      \end{align*}
\end{restatable}

As one can see from the diagrams (\ref{3jstatedgm}) and (\ref{injectiondgm}), these $3jm$-symbol have more symmetries than the Clebsch-Gordan coefficients because they treat the three Hilbert spaces on the same level.
We can now diagrammatically prove the symmetry properties of the $3jm$-symbol.

\begin{proposition}\label{prop:threejsymbolevenswap}
  A 3-j symbol is invariant under an even permutation of its columns:
  \begin{align*}
    \left(
      \begin{array}{ccc}
        j_1 & j_2 & j_3 \\
        m_1 & m_2 & m_3
      \end{array}
    \right)
    \quad&=\quad
    \left(
      \begin{array}{ccc}
        j_2 & j_3 & j_1 \\
        m_2 & m_3 & m_1
      \end{array}
    \right)\\
    \quad&=\quad
    \left(
      \begin{array}{ccc}
        j_3 & j_1 & j_2 \\
        m_3 & m_1 & m_2
      \end{array}
    \right)
  \end{align*}
  Diagrammatically, this is stated as follows:
  \[
    \scalebox{.75}{\input{figures/3jsymbolcalc/3jmevenpermute-alt-lemmav2-2col.tikz}}
  \]
  where permutation of the columns of a 3j-symbol is realised by dragging the corresponding parts of the diagram to the right place.
\end{proposition}
\noindent The following property can be proved similarly, by invoking the last equality of \cref{eq:simp-lemma-1-pf}.
\begin{proposition}\label{prop:threejsymboloddswap}
  An odd permutation of the columns gives a phase factor:
  \begin{align*}
    \begin{array}{ll}
      \left( \begin{array}{ccc}
               j_1 & j_2 & j_3 \\
               m_1 & m_2 & m_3
      \end{array} \right) &=(-1)^{j_1 +j_2 + j_3} \left( \begin{array}{ccc}
              j_2 & j_1 & j_3 \\
              m_2 & m_1 & m_3
      \end{array} \right) \vspace{0.3cm} \\
      &=(-1)^{j_1 +j_2 + j_3}
      \left( \begin{array}{ccc}
               j_1 & j_3 & j_2 \\
               m_1 & m_3 & m_2
      \end{array} \right) \vspace{0.3cm} \\
      &=(-1)^{j_1 +j_2 + j_3}
      \left( \begin{array}{ccc}
               j_3 & j_2 & j_1 \\
               m_3 & m_2 & m_1
      \end{array} \right)
    \end{array}
  \end{align*}
\end{proposition}
\noindent Furthermore, we have
\begin{restatable}{proposition}{threejsymbolsignflip}
  Changing the sign of the
  $m$ quantum numbers gives a phase:
  \[
    \left(\!
    \begin{array}{ccc}
             j_1  & j_2  & j_3  \\
             -m_1 & -m_2 & -m_3
    \end{array}
    \!\right)
    \!=\!
    (-1)^{j_1 +j_2 + j_3} \left(\!
    \begin{array}{ccc}
      j_1 & j_2 & j_3 \\
      m_1 & m_2 & m_3
    \end{array}
    \!\right)
  \]
  Diagrammatically, this is stated as follows:
  \[
    \scalebox{.75}{\input{figures/3jsymbolcalc/3jmvaluepermute-alt-lemmav2-2col.tikz}}
  \]
\end{restatable}

\section{The Lie algebra \texorpdfstring{$\mathfrak{su}(2)$}{su(2)}}\label{sec:lie-algebra-su2}

One immediate advantage of the embedding is that it expands the scope of what the Spin-ZX calculus as diagrams can formally interact with. In particular, we can straightforwardly describe via diagrams properties of the related Lie algebra $\mathfrak{su}(2)$. As all the elements are embedded into ZX it is simpler to formally interact elements of the algebra and group while maintaining confidence in the formal meaning of what is depicted and calculated.

We first introduce the angular momentum operators. The angular momentum operators are the generators of $SU(2)$ in the sense that they are a basis for the group's Lie algebra $\mathfrak{su}(2)$. Any $g\in SU(2)$ can be represented as the exponent of some $\mathfrak{g} \in \mathfrak{su}(2)$, specifically $g = e^{i\mathfrak{g}}$. The elements of $SU(2)$ and $\mathfrak{su}(2)$ are also in local correspondence, as the derivate of the group at the identity gives the Lie algebra. In this way, the algebra elements also act as infinitesimal transformations (i.e.\@ generators of $SU(2)$).

They can be defined in terms of the magnetic basis~\footnote{The $m$ can be seen as the alignment of the angular momentum in the z-direction (an arbitrary choice but one used by convention), along which one could discriminate these states physically by applying a magnetic field. The variation in the energy levels due to doing this is the famous Zeeman effect.} as follows:
\begin{align*}
    J_1\ket{j;m} = &\phantom{+} \frac{1}{2}\sqrt{(j-m)(j+m+1)}\ket{j;m+1}\\
    &+ \frac{1}{2}\sqrt{(j+m)(j-m+1)}\ket{j;m-1} \vspace{0.3cm}\\
    J_2\ket{j;m} = &\phantom{+} \frac{1}{2i}\sqrt{(j-m)(j+m+1)}\ket{j;m+1}\\
    &- \frac{1}{2i}\sqrt{(j+m)(j-m+1)}\ket{j;m-1} \vspace{0.3cm}\\
    J_3\ket{j;m} = &\phantom{+} m\ket{j;m}
\end{align*}
The matrices $D^j_g$ representing the operators $g = e^{-i\alpha\overrightarrow{n}\cdot \overrightarrow{J}}$ on the Hilbert space $H_j$  in the $\ket{j;m}$ basis define a 
$(2j + 1)$-dimensional irreducible representation of SU(2), where $\overrightarrow{J}=(J_1, J_2, J_3)$~\cite{makinenIntroductionSU22019}. These matrices $D^j_g$,  called \emph{Wigner matrices}, are square matrices of size $2j + 1$, whose elements are given by~\cite{makinenIntroductionSU22019, martin-dussaudPrimerGroupTheory2019}:
\[
D^j{mn}(g) = \bra{j;m} e^{-i\alpha\overrightarrow{n}\cdot \overrightarrow{J}}\ket{j;n}
\]

The angular momentum operators $J_i$ generate a non-abelian rotation group.
They are characterised by the \emph{fundamental commutation relations of angular momentum}:
\begin{equation}\label{angularmomentumcommutation}
  [J_j, J_k] = i\, \epsilon_{jkl} J_l
\end{equation}
where the Levi-Civita symbol  $\epsilon_{jkl}$ is $1$ if $(j, k, l)$ is an even permutation of $(1, 2, 3)$, $-1$ if it is an odd permutation, and $0$ otherwise. Alongside these generators, the other key operators of interest are the ladder operators which allow us to transition between the states within a representation
\[
  J_{+} \coloneqq J_1 + iJ_2 \quad \text{and} \quad J_{-} \coloneqq J_1 - iJ_2
\]
whose action on the magnetic basis is
\begin{equation*}
  \begin{array}{l}
    J_+\ket{j;m} = \sqrt{(j-m)(j+m+1)}\ket{j;m+1} \vspace{0.3cm}\\
    J_-\ket{j;m} = \sqrt{(j+m)(j-m+1)}\ket{j;m-1}.
  \end{array}
\end{equation*}
Using the ladder operators, we can write the commutation relations equivalent to~\eqref{angularmomentumcommutation} as:
\begin{equation}\label{equivalentcom}
[J_+, J_-] = 2J_3 \quad \text{and} \quad [J_3, J_{\pm}] = \pm J_{\pm}.
\end{equation}
It is easy to check that the ladder operators can be expressed as the following diagrams.
\[
  \input{figures/joperator/joperators.tikz}
\]
where
\begin{align*}
L_j = & 1\frac{1}{\sqrt{2j}} \big(\sqrt{2(2j-1)}, \cdots, \sqrt{(k+1)(2j-k)}, \cdots,\\
&\sqrt{(2j+1)(2j-2j)}\big).
\end{align*}
We can add or subtract the above diagrams to construct the angular momentum operators $J_1$ and $J_2$.
\[
  \input{figures/joperator/j1operators-2col.tikz}
\]
This allows us to define
\[
  \input{figures/joperator/j1operators2-1.tikz}
\]
and
\[
  \input{figures/joperator/j1operators2-2.tikz}.
\]
Finally, to obtain the diagram for $J_3$, we observe that it is a diagonal operator given by
\[
  \input{figures/joperator/j3operator.tikz}
  \text{   where    } A_j = \frac{1}{j} (j-1, \ccdots, j-k, \ccdots, j-2j).
\]

Next, we derive the commutation relations given in~\eqref{equivalentcom}.
\begin{restatable}{proposition}{jpjmcommute}
  $[J_+, J_-] = 2J_3$, or diagrammatically,
  \begin{equation}
    \input{figures/joperator/j+j-commute-prop.tikz}
  \end{equation}
\end{restatable}

\begin{restatable}{proposition}{jthreejpcommute}
  $[J_3, J_{+}] = J_{+}$, or diagrammatically,
  \begin{equation}
    \input{figures/joperator/j3j+commute-prop.tikz}
  \end{equation}
\end{restatable}

\begin{restatable}{proposition}{jthreejmcommute}
  $[J_3, J_{-}] = - J_{-}$, or diagrammatically,
  \begin{equation}
    \input{figures/joperator/j3j-commute-prop.tikz}
  \end{equation}
\end{restatable}

\subsection{Expressing Hamiltonians}

The ZXW calculus has previously been used to sum and exponentiate qubit Hamiltonians~\cite{shaikhHowSum2023}.
Similarly, it is possible to extend our generators, enabling the Spin-ZX calculus to express Hamiltonians in terms of angular momentum operators provided we also make use of the ZXW W node
\[
\input{qufit-generators/w1to2.tikz}
\quad \overset{\interp{\cdot}}{\longmapsto} \quad
\ket{00}\bra{0}+\sum_{i=1}^{d-1}(\ket{0i}+\ket{i0})\bra{i}.
\]
and the triangle node
\[
\input{qufit-generators/triangled.tikz}
\quad \overset{\interp{\cdot}}{\longmapsto} \quad
I_d+\sum_{i=1}^{d-1}\ket{i}\bra{0}.
\]

We can then express the controlled $J_1$, $J_2$, and $J_3$ operators in our calculus respectively as follows:
\begin{align*}
    \input{figures/joperator/controlled-J1.tikz}\\
    \input{figures/joperator/controlled-J2.tikz}\\
    \input{figures/joperator/controlled-J3.tikz}
\end{align*}
where $d=2j+1$ and $\overrightarrow{0}\!,\!\overrightarrow{1} = (\overbrace{0, \cdots, 0}^{d - 1}, \overbrace{1, \cdots, 1}^{d - 1})$.


Then a Hamiltonian composed of angular momentum operators
 can be written as a self-adjoint operator of the form:
\begin{equation}
 H = \sum_{r=1}^m \alpha_{r} \bigotimes_{i = 1}^n J_{ri}^{k_{ri}} 
\end{equation}
where each $ \alpha_{r} $ is a real number, $J_{ri} \in \{J_1, J_2, J_3\}$. 
 It can be diagrammatically represented as follows:
\begin{widetext}
  \begin{equation}
      \input{figures/joperator/hamiltonian.tikz}
  \end{equation}
\end{widetext}
We can now 
exponentiate the Hamiltonian as done in~\cite{shaikhHowSum2023} as follows:
\begin{equation}
  \input{figures/joperator/hamiltonian-exponential.tikz}  
\end{equation}
for some coefficients $c_0,\dots,c_k$ where $k$ is the dimension of the Hilbert space.

\section{Applications}\label{sec:app}
\subsection{Permutational quantum computing}\label{pqc}

\begin{figure*}[ht]
	\[
		\input{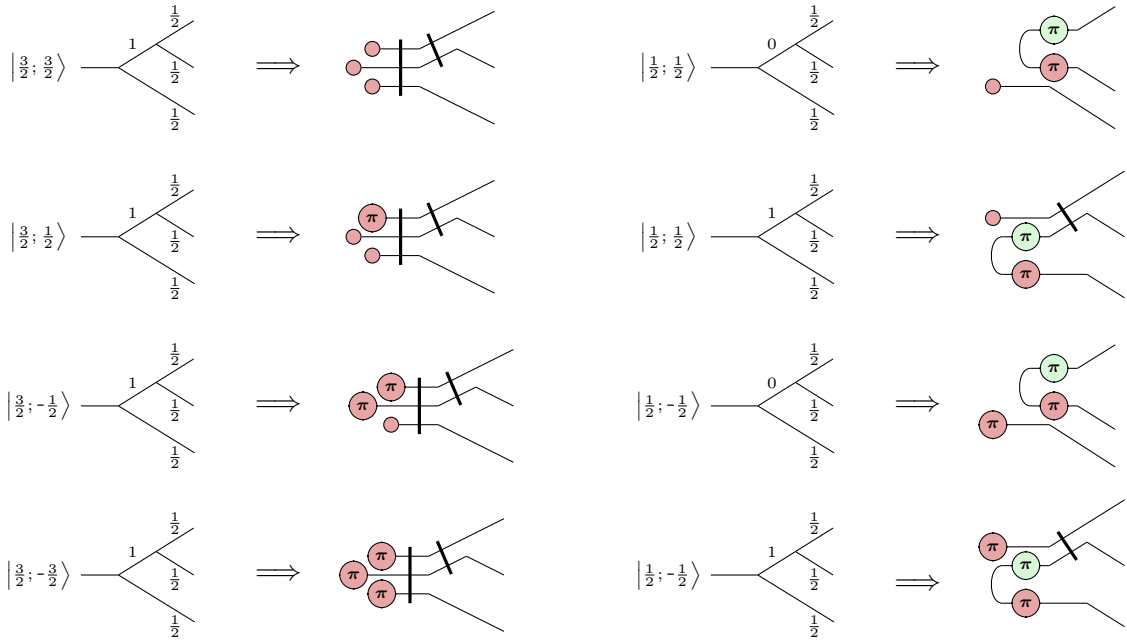}
	\]
	\caption{Graphical depiction of a coupling basis of three qubits, where the pairwise coupling of the spaces proceeds from the left (other possibilities give alternative bases).
	The top row details the ways one can obtain total angular momentum of $\frac{1}{2}$, the second of  $\frac{3}{2}$.
	The elements in the rows correspond to the specific different states, giving a final $m$ value on the spaces at the bottom of the trees.
	Note that in the absence of specifying the $m$ values, the set of diagrams belongs to one of a class of three different spin networks determined by the recoupling outcomes in the tree.}
	\label{fig:trees}
\end{figure*}

An early model of spin-network computation was introduced by Marzuoli and Rasetti in~\cite{marzuoliComputingSpin2005}.
This, in turn, was specialised into a formal computational model called permutational quantum computing (PQC) by Jordan in~\cite{jordanPermutationalQuantum2010} by restricting computations to the expectation values of permutations between states defined by binary-tree-shaped spin networks that can be more simply understood as the basis of total angular momentum outcomes from a collection of qubits.
Consider three qubits in the standard computational basis $\{b_i b_j b_k\}$ where each $b$ is a binary value detailing if the qubit is spin up or spin down.
It can be shown that an alternative \emph{spin basis} is given by pairwise coupling via total angular momentum measurements.
For instance, we could measure the angular momenta of the first two qubits, getting either spin-1 or spin-0 and then getting the total angular momentum measurement for the spin of this outcome with the final qubit leading to total spin-$\frac{1}{2}$ or spin-$\frac{3}{2}$.
If we then also fix the state of this final angular momentum's $m$-value, i.e.\@ its z-direction angular momentum, this also gives us a basis (and indeed, any other choice of a fixed coupling pattern is another such basis: see \cref{fig:trees} for an example).

\begin{figure}[ht]
	\[
		\input{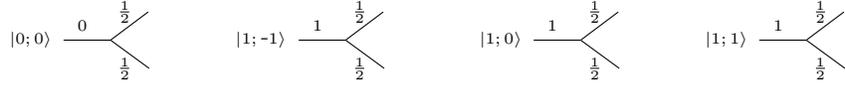}
	\]
	\caption{A graphical depiction of how the basis constructed by combining angular momentum of two spin-$\frac{1}{2}$ systems, and the possible outcomes of total and $z$-directed angular momenta.}
	\label{fig:spin-recup}
\end{figure}

The PQC model is the class of computations defined as expectation values of qubit swap operators on states prepared in (one of the) spin bases.
These computations have previously been done by hand by making use of the Racah identities, and the interested reader is directed to~\cite{jordanPermutationalQuantum2010,marzuoliComputingSpin2005} for further explanation and details.
Here, however, by using the work in~\cite{eastSpinnetworksZXcalculus2022} it is possible to simply represent these computations as ZX-based tensor networks and with the symmetry-based rewrite language described here reduce these down to scalar values.

Consider the following example where the final leaves of the tree are all spin-$\frac{1}{2}$ i.e. qubits:
\begin{equation}
	\scalebox{0.75}{\input{figures/qufit-applications/Jordon-ex-init.tikz}}
	\quad=\quad
	\frac{\sqrt{3}}{2}
\end{equation}
which was calculated in~\cite{jordanPermutationalQuantum2010} via the Racah transformation rules in the manner first seen in~\cite{marzuoliComputingSpin2005}.
To alternatively represent this in a directly graphical manner, we first find the normalisations of the PQC spin basis states as diagrams.
For the basis element where the first coupling forms spin-1, we find
\begin{align*}
    \hspace*{1cm}&\hspace{-1cm}
		\scalebox{0.75}{\input{figures/qufit-applications/pqc-jordon-norm-0.tikz}} \vspace{0.3em}\\
	&\scalebox{0.75}{\input{figures/qufit-applications/pqc-jordon-norm-1v2.tikz}} \vspace{0.3em}\\
	&\scalebox{0.75}{\input{figures/qufit-applications/pqc-jordon-norm-2v2-1.tikz}} \vspace{0.3em}\\
	&\scalebox{0.75}{\input{figures/qufit-applications/pqc-jordon-norm-2v2-2.tikz}}
\end{align*}
For the element where they form a spin-0, we find that
\begin{equation}
	\scalebox{0.75}{\input{figures/qufit-applications/pqc-jordon-norm1v2-2col.tikz}}
\end{equation}

We can then represent the whole computation as a closed diagram composed of the representations of the states with a permutation between them.
The solution is given by applying the rewrite rules:

\begin{equation}
	\scalebox{0.75}{\input{figures/qufit-applications/pqc-zxw-examp-jordonv2-2col.tikz}}
\end{equation}

Where the initial scalar of $\frac{1}{\sqrt{3}}$ is the overall normalisation given the inner product of the two initial states.

\subsection{Analysis of SPT phases}

In Ref.~\cite{eastAKLTStatesZXDiagramsDiagrammatic2022} it was shown that a modification of the ZX calculus could find use beyond quantum computing and in broader theoretical physics.
In particular, it was used to analyse the AKLT (Affleck-Kennedy-Lieb-Tasaki) state.
The form presented in~\cite{eastAKLTStatesZXDiagramsDiagrammatic2022} contains a large degree of redundancy due to the true focus of the calculus used being qubits quantum computing and not spin, in this section we will show some of the same arguments but in a simpler fashion with simpler derivations.
The AKLT state~\cite{affleckRigorousResultsValencebond1987, affleckValenceBondGround1988}, named after Affleck, Lieb, Kennedy, and Tasaki, is a quantum state that arises as the ground state of the one-dimensional AKLT Hamiltonian.
This Hamiltonian is defined for a chain of spin-1 degrees of freedom, where each `site' (element in the chain) interacts with its neighbour in the following manner:

\[ H = \sum_i \left( \vec{S}_i \cdot \vec{S}_{i+1} + \frac{1}{3}(\vec{S}_i \cdot \vec{S}_{i+1})^2 \right) \]

Here, \(\vec{S}_i\) represents the spin-1 operator at site \(i\), and the sum runs over all pairs of neighbouring sites.
The AKLT state is formed of spin-1 sites partitioned into two spin-1/2 sites.
These spin-1/2 sites then form singlet states with the neighbouring sites, creating a state with a maximum total spin of 1 between neighbouring spin-1's.
This ensures the state is annihilated by the spin-2 projectors included in the Hamiltonian to give us the ground state.

For our interests, the AKLT state has two notable properties:
\begin{enumerate}
	\item When the chain has open boundary conditions, the termination of the chain leaves unpaired spin-1/2 particles at each end, leading to a four-fold degeneracy due to the two possible states (spin up or down) of each edge spin.
	\item The AKLT state exhibits a non-local hidden order known as string order.
	\item This order is characterised by being an equal-weight superposition of antiferromagnetic states where, ignoring the zero spin states, the remaining spins alternate in orientation (anti-ferromagnetic order).
	\item This means that a configuration like \(|1, 0, 0, 0, -1, 0, 0, 0, 1\rangle\) is allowed, while \(|1, 0, 0, 1, 0, 0, 0, 1\rangle\) is not.
\end{enumerate}

Proceeding in the manner of Ref.~\cite{eastAKLTStatesZXDiagramsDiagrammatic2022}, we connect the AKLT state's MPS (matrix product state) representation to a diagrammatic one.
Recalling that any quantum state can be written as a product of matrices, i.e.\@ as a matrix product state, via
\begin{equation}
	|\psi\rangle\ =\ \sum_{\substack{j_1, \ccdots, j_N\\ \alpha_2, \ccdots, \alpha_{N+1}}} M^{[1]j_1}{}_{\alpha_1,\alpha_2} \ccdots M^{[N]j_N}_{\alpha_N,\alpha_{N+1}}\ket{j_1, j_2, \ccdots, j_N}.
\end{equation}

The indices $j_i$ are called physical indices because they span the local Hilbert space at a given site $n$ for example, $j_i = 0, \pm1$ for spin-1.
For a given $j_i$ and $n$, the $M^{[n]j_n}_{\alpha_i,\alpha_{i+1}}$ are matrices in the indices $\alpha_i$, known as bond indices.
Though this does offer an exact representation of any finite system, the maximum dimension of the bond indices, or the bond dimension $\chi$, typically grows exponentially with system size.

The AKLT state is notable for its simple representation as an MPS with bond dimension $\chi = 2$ meaning each site is defined by a $2\times 2$ matrix (with a $1\times 2$ matrix at the boundaries).
As each site in the AKLT state is spin-1 we can see that the AKLT state as an MPS is defined by three matrices
\begin{align}
	M^{[n]}_{+1} = \sqrt{\frac{2}{3}}
	\begin{pmatrix} 0 & 0 \\ 1 & 0 \end{pmatrix}, \qquad & \qquad
	M^{[n]}_{0} = \frac{1}{\sqrt{3}}
	\begin{pmatrix} 1 & 0 \\ 0 & -1 \end{pmatrix}, \nonumber\\
	M^{[n]}_{-1} &= \sqrt{\frac{2}{3}}
	\begin{pmatrix} 0 & -1 \\ 0 & 0 \end{pmatrix}.
\end{align}
which are the same for all sites $1 < n < N$ in the bulk.
In the Spin-ZX calculus, we can see these as particular spin states applied to the spin-1 projectors:
\begin{align*}
	&\scalebox{.75}{\input{figures/qufit-applications/AKLT-red-1-2col.tikz}}\\
	&\scalebox{.75}{\input{figures/qufit-applications/AKLT-red-2-2col.tikz}}\\
	&\scalebox{.75}{\input{figures/qufit-applications/AKLT-red-3-2col.tikz}}
\end{align*}
The general chain can therefore be represented as described above as singlets linked by spin-1 projectors.
\begin{equation}
	\scalebox{0.75}{\input{figures/qufit-applications/AKLT.tikz}}
\end{equation}

Furthermore, we can demonstrate that the AKLT state is a ground state of the AKLT Hamiltonian.
To see this we first rewrite the AKLT Hamitonian as a sum of Spin-2 projectors (see appendix A of~\cite{eastAKLTStatesZXDiagramsDiagrammatic2022} for details):
\begin{align}
	H &= \sum_{i} \vec{J}_i \cdot\vec{S}_{i+1} + \frac{1}{3}  (\vec{S}_i \cdot\vec{S}_{i+1})^2 \nonumber\\
	&= 2\sum_{i} \left(P^{(2)}(\vec{S}_{i},\vec{S}_{i+1})- 1/3\right).
	\label{eq:AKLThamapp}
\end{align}

Where $P^{(2)}$ is the projector on two sites to the spin-2 subspace. The spin-2 projector being positive or zero we can see that for every two sites in a spin-2 state, the energy rises. The corollary is that the ground state must be a chain where no pair of sites is in the spin-2 state.
	
Consider now then that for any two sites in our diagrammatic AKLT state, the following gives zero.

\begin{proposition}
	\[
		\scalebox{.75}{\input{figures/qufit-applications/aklt-ground-state.tikz}}
	\]
\end{proposition}
\begin{proof}
	\begin{equation*}
		\scalebox{.75}{\input{figures/qufit-applications/aklt-ground-state-proof-2col.tikz}}
	\end{equation*}
\end{proof}
As the operator acting on each pair of sites is the map to the symmetric sub-space of 4 2D spin-$\frac{1}{2}$ spaces then we can know this is a projector on two spin-2. In turn, this indicates any pair of sites in the chain are sent to 0 by the spin-2 projector which indicates the AKLT groundstate.

It was also shown in Ref.~\cite{eastAKLTStatesZXDiagramsDiagrammatic2022} that one can investigate AKLT order diagrammatically.
They did this by diagrammatically demonstrating that only Anti-ferromagnetic choices for the sites give non-zero scalars. The Spin-ZX calculus allows for a simpler version of the diagrammatic proof given there. Firstly, we see $\ket{j=1,m=0}$ sites result in bell state links when simplified:
\begin{equation}
	\scalebox{0.75}{\input{figures/qufit-applications/chain-red-2col.tikz}}
\end{equation}
We can then see that no matter the number of intervening $\ket{j=1,m=0}$ sites only sites with opposing spins are allowed in sequence ($\ket{j=1,m=1}$ then $\ket{j=1,m=-1}$ or vice versa).
\begin{align}
	&\scalebox{0.75}{\input{figures/qufit-applications/00-red.tikz}}\;\;=\;\;0\;, \nonumber \\
	&\scalebox{0.75}{\input{figures/qufit-applications/10-red.tikz}}.
\end{align}

Another feature of the diagrammatic representation of states is that it is simple to represent higher dimensional AKLT states which are formed from n-qubit symmetrisers acting on one end of n singlets projected to spin-$\frac{n}{2}$.
For example, the 2D AKLT honeycomb lattice is simply:
\begin{restatable}{proposition}{hexagontwodaklt}
We can transform from the Penrose-style diagram to the Yutsis-style diagram by composing the former with isometries~\eqref{eq:isometry} and simplifying it to get the Honeycomb shape.
	\begin{align}\label{2daklt}
		&\scalebox{0.75}{\input{figures/qufit-applications/2daklt.tikz}}
		\nonumber \\
		\rightsquigarrow \qquad
		&\scalebox{0.75}{\input{figures/qufit-applications/2daklt-hexagon.tikz}}
	\end{align}
\end{restatable}

In Ref.~\cite{eastAKLTStatesZXDiagramsDiagrammatic2022} this lattice was used to show how the random outcomes of single-site POVM measurements reduce the state to a graph state, and hence that it is a universal resource for measurement-based quantum computation.
Each honeycomb site is a spin-$\tfrac{3}{2}$ degree of freedom, given by the symmetric subspace of three qubits, one for each singlet bond, and is measured by a three-element POVM.
Each outcome projects these three qubits onto a two-dimensional subspace, i.e.\ a single encoded qubit, giving
\begin{equation}
	\label{eq:2d-red}
	\scalebox{0.75}{\input{figures/qufit-applications/2d-red-2col.tikz}}
\end{equation}
where $\vec{a}=\left(1, \frac{1}{3},\frac{1}{3}\right)$ and each rewrite is labelled with the rule used.
The real-basis ($Z$ and $X$) outcomes give
\begin{equation}
	\scalebox{0.75}{\input{figures/qufit-applications/2d-realred-2col.tikz}}
\end{equation}
while the remaining outcome carries a relative $\pi/2$ phase,
\begin{equation}
	\scalebox{0.75}{\input{figures/qufit-applications/2d-redpiby2-2col.tikz}}
\end{equation}
Applying these at each site and simplifying, every measured site becomes a single graph-state vertex, with the random outcomes fixing only local Clifford corrections.
The lattice in \cref{2daklt} thus reduces to a graph state, though as compared with the original paper the demonstration of this is more diagrammatically elegant (cf.\ appendix~2 in Ref.~\cite{eastAKLTStatesZXDiagramsDiagrammatic2022}).

In the original work, further properties were demonstrated diagrammatically which can also be converted to the Spin-ZX calculus; however, it is not our intention to replicate all the results of that paper.
The ambition is to highlight to interested readers from the domain of condensed matter physics that the Spin-ZX calculus is a versatile tool for $SU(2)$-based system in this domain.

\subsection{Analysis of QML ansätze}

The ZX calculus can be used to represent derivatives and integrals of maps and states represented as diagrams themselves~\cite{zhaoAnalyzingBarren2021,wangDifferentiatingIntegrating2024}. One application of this is quantum machine learning.

Here we make use of the spin-network ansatz seen in Ref.~\cite{eastAllYou2023} which uses parametrised $SU(2)$ equivariant unitaries to learn maps and states which possess this symmetry. Let us focus on the two-qubit unitary vertex gate of Ref.~\cite{eastAllYou2023}.
It functions by taking the fact that $\frac{1}{2}\otimes\frac{1}{2}\simeq0\oplus 1$ and applying a phase to the spin-0 component.
Essentially constructing the map $V(\theta) = \ket{J=1}\bra{J=1}+e^{i\theta}\ket{J=0}\bra{J=0}$ which we can write as a sum of projectors with a phase as:

\begin{equation}\label{eq:2q-vert-gate}
	\input{figures/qufit-applications/vert-gate-2col.tikz}
\end{equation}

To reach a single diagram form we make use of the fact that the above is a direct partition of the qubit space into the block diagonal spin basis. The quantum gate that achieves this is the 2-qubit Schur transform:
\[
	\input{figures/qufit-applications/schurgate-2col.tikz}
\]
Here, we used a shorthand for the controlled-Hadamard gate, defined as follows:
\[
	\input{figures/qufit-applications/controlhadamardeq.tikz}
\]
Following~\cite{eastAllYou2023}, we can now express $V(\theta)$ using these definitions as a single diagram:
\begin{equation}\label{vert-gate-uni}
	\input{figures/qufit-applications/vert-gate-uni2-2col.tikz}
\end{equation}
where $P_2(\theta)$ is the controlled $Z$-rotation, defined as follows:
\[
	\input{figures/qufit-applications/p2gate.tikz}
\]
where
\[
	\input{figures/qufit-applications/vert-gate-and.tikz}
\]
This applies the phase to the spin-0 sector once it is mapped to a single qubit computational basis element by the Schur transform.

As an application of this ansatz and calculus on it consider the permutational quantum computing model described above (\cref{pqc}). Here the expectation of the Hamiltonian is written as:
\[
	\langle H\rangle ~=~ \bra{\psi} W(\vec\theta) S_n W^\dagger(\vec\theta) \ket{\psi}
\]
where $W(\vec\theta)$ is a brick-wall ansatz made up of a tiling of the vertex gates $V(\theta)$:
\[
	\input{qufit-applications/brickwork-ansatz.tikz}
\]
and the PQC Hamiltonian itself $S_n$ is an arbitrary permutation of $n$ qubits, with the initial states $\ket{\psi}$ set as the product of singlets, i.e.\@ the spin-0 state:
\[
	\ket{\psi} = \left(\input{qufit-applications/y-cap.tikz}\right)^{\otimes k}
\]
that can be physically implemented as
\[
	\input{qufit-applications/making-y-cap.tikz}
\]

Given that each parameter appears exactly once in the ansatz, we can represent the expectation values as follows~\cite{kochContractionZX2024}:
\begin{align*}
	\langle H\rangle
	~&=~ \input{qufit-applications/exp-1.tikz}\\
	&\coloneqq~ \input{qufit-applications/exp-2.tikz}\\
	~&=~ \bra{0}U^\dagger(\vec\theta)HU(\vec\theta)\ket{0}
\end{align*}
Since $H$ can be represented as a diagram, we can calculate the gradient variance as follows~\cite{wangDifferentiatingIntegrating2024}:
\begin{widetext}
\begin{equation} \label{eq:variance}
	\operatorname{Var}\left(\frac{\partial\langle H\rangle}{\partial \theta_j}\right) ~=~ \input{qufit-applications/variance.tikz}
\end{equation}
\end{widetext}
Such a diagram can then be simulated using standard ZX-based techniques, such as the one given in Ref.~\cite{kochContractionZX2024}.

\subsection{Quantised volume calculations in loop quantum gravity}

Though a full introduction to loop quantum gravity (LQG) is beyond the scope of this paper, the key elements for our purposes are that it proposes that space itself is quantised and that properties like area and volume can be obtained from the appropriate area and volume operators.
In LQG the Hilbert space of space itself is given by the sum over all possible 4-valent spin networks.
Each particular spin network can be viewed as a geometric object by looking at its dual, which gives us an image of tessellating tetrahedra, each edge corresponding to a face, each intertwiner corresponding to a volume, with intertwiners sharing edges indicating a shared face of two tetrahedra.
These can all be represented in ZX form as was first discussed in Ref.~\cite{eastPhysiqueFormelleSpin2022}.
Given this perspective, LQG tells us that the area of a face is given simply by the operator:
\begin{equation}
	\hat{A}_l\ =\ 8\pi\gamma\frac{\hbar G}{c^3}\sqrt{\vec{J}_l^2}
\end{equation}
which has the following eigenstates:
\begin{equation}
	\hat{A}_l\left|\Gamma\right\rangle\ =\ 8\pi\gamma\frac{\hbar G}{c^3}\sqrt{j_l(j_l+1)}\left| \Gamma\right\rangle
\end{equation}
We can see that, up to some fundamental constants, the area operator for a face of a quantised tetrahedron (given by a spin-network's edge) is directly proportional to the magnitude of the total angular momentum operator.

The case of volume is more challenging as the operator is not diagonal in this basis.
The volume operator is given by:
\begin{equation}
	\hat{V}\ =\ \frac{\sqrt{2}}V(8\pi G\hbar\gamma)^{3/2}\sqrt{|\vec{J}_1\cdot(\vec{J}_2\times\vec{J}_3)|}\label{vol}
\end{equation}
We can see the core calculational element here can be isolated as:
\begin{equation}
	V^2\ \propto\ \tilde{V}^2\ =\ \epsilon_{ijk}J_x^iJ_y^jJ_z^k\ =\ \frac{1}{8}\epsilon_{ijk}\sigma_x^i\sigma_y^j\sigma_z^k
\end{equation}
We can construct this operation in the Spin-ZX calculus by making use of dimension splitters:
\begin{equation}
	 \input{qufit-axioms/express-dimension-split.tikz}
\end{equation}
and writing the controlled versions of the Pauli matrices as follows:
\begin{equation}
	\input{figures/qufit-applications/controlled-pauli.tikz}
\end{equation}
where $P_i$ for $i = 0, 1, 2, 3$ is $I$, $\sigma_x, \sigma_y$ and $\sigma_z$ respectively.
With this, we construct $\tilde{V}^2$:
\begin{equation}
	\tilde{V}^2
	\quad=\quad
	 \frac{i}{8}\scalebox{0.75}{\input{figures/qufit-applications/vol-opv2.tikz}  }
\end{equation}
where the yellow box is the usual Hadamard gate.
An equivalent of this diagram was used in ZXH to demonstrate that the state corresponded to the minimum spin-$\frac{1}{2}$ volume.
It can be shown that the eigenstate of this volume is given by the sum $\iota_0+i\sqrt{3}\iota_1$, where $\iota_0$ and $\iota_1$ are the respective recoupling possibilities for two spin-$\frac{1}{2}$, namely spin-0 and spin-1 (see~\cite{martin-dussaudPrimerGroupTheory2019} for more details).
Diagrammatically, we can represent this as follows:
\begin{align}
	\left|\iota_0+i\sqrt{3}\iota_1\right\rangle
	\quad&=\quad
	\frac{i}{\sqrt{3}}\begin{pmatrix}
		0 & 0 & 0 & 1\\
		0 & -e^{i\frac{\pi}{3}} & e^{i\frac{2\pi}{3}} & 0\\
		0 & e^{i\frac{2\pi}{3}} & -e^{i\frac{\pi}{3}} & 0\\
		1 & 0 & 0 & 0
	\end{pmatrix} \nonumber \\
	&=\quad \frac{1-\frac{i}{\sqrt{3}}}{4}\scalebox{0.75}{\input{figures/qufit-applications/vol-min-intv2.tikz}}
\end{align}
where we use a slight abuse of notation by denoting this map as a state. The question then becomes about witnessing the action of this operator on the volume to see if is an eigenvector and retrieve the eigenvalue. We can see this for the minimal 4 spin-1/2 volume as follows
\begin{restatable}{proposition}{volumeoperator}
The state $\left|\iota_0+i\sqrt{3}\iota_1\right\rangle$ is  an eigenvector of the operator $I \otimes \tilde{V}^2$ with eigenvalue $ \frac{-\sqrt{3}}{4}$.
\begin{align}
	 \scalebox{0.75}{\input{figures/qufit-applications/vol-state.tikz}}
	 \nonumber \\
	 =\quad
	\frac{-\sqrt{3}}{4}\scalebox{0.75}{\input{figures/qufit-applications/vol-min-intv3.tikz}}
\end{align}
\end{restatable}
Comparing with the literature~\cite{rovelliCovariantLoop2014} we see we indeed obtain the correct minimal volume (recalling from equation \eqref{vol} that we are only interested in the modulus so the volume is not imaginary). This calculation is of broader interest being one of the first examples of a diagrammatic eigenvector calculation. The determination of the eigenvectors of diagrammatic operators is a broad question and one that, though unaddressed further here, merits further study.

\section{Conclusion and discussion}\label{sec:future}
In this paper, we have introduced the Spin-ZX calculus as a formal diagrammatic language tailored for the representation theory of $SU(2)$, elevating Penrose's original diagrams into a full language for quantum information and spin physics. Our work demonstrates the versatility of the Spin-ZX calculus across multiple domains. In permutational quantum computing, we have demonstrated how transition amplitudes can be derived entirely through diagrammatic reasoning, simplifying calculations that traditionally required a knowledge of Racah transforms. In condensed matter physics, the calculus offers a more natural and streamlined approach to analysing AKLT states, extending beyond previous qubit-based diagrammatic techniques. In quantum machine learning, we have illustrated how parametrised quantum circuits for spin systems can be analysed, providing insights into the variance of $SU(2)$ equivariant ansätze. In quantum gravity, our techniques simplify the representation of the volume operator in loop quantum gravity, enabling the calculation of minimal quantised volumes through purely diagrammatic methods.

The Spin-ZX calculus not only simplifies the representation and manipulation of $SU(2)$ systems but also offers new ways to look at the foundational relationships within representation theory. By providing a formal diagrammatic language, we pave the way for the development of new algorithms and methodologies in theoretical physics. This work represents a significant step towards integrating quantum information science with the fundamental symmetries of nature, and we anticipate that it will inspire further research and applications in both quantum computing and broader theoretical physics. Future work will continue the effort to diagrammatise physics by the creation of calculi for representations of $SU(N)$, in particular $SU(3)$, which will allow for the natural treatment of systems of interest to particle physicists and allow for the diagrammatic description of the standard model. It will also look to apply the Spin-ZX calculus to more $SU(2)$ systems such as more 2D lattices, spin glasses, $SU(2)$ resistant decoherent free subspaces, and error correction. In addition it will seek to create new methods for devising algorithms of interest to physicists and computer scientists alike by leveraging this new perspective on quantum information in spin systems.

In general we note that quantum information is something of a curiosity in terms of foundational physics. In other domains of physics pertaining to the basic structure of nature, symmetries reign supreme -- indeed the definitions of what things in themselves \emph{are} tend to be expressed in terms of symmetries. Particles from photons and electrons to quarks and anyons are all tied to various (representations of) symmetry groups. 
In the case of quantum information however the dominant theoretical framework is registers of qubits, which are ultimately an engineering principle.
We often want quantum algorithms to analyse natural physical systems in which symmetry groups are the characteristic feature. 
As such, we should devise tools where the presentation of these symmetries is not just possible but natural.

\newpage
\appendix

\section{Traditional Penrose diagrams and ZX}
\label{sec:penrose_diagrams}

Penrose diagrams (as opposed to the Spin-ZX calculus we introduce here) are a diagrammatic notation introduced by Roger Penrose to reason about spins.
A form of these diagrams decomposed into strands of spin-$\frac{1}{2}$ i.e.\@ qubits is of particular interest to us - often termed the binor calculus~\cite{martin-dussaudPrimerGroupTheory2019}.
It is this that we alter and raise to the level of a diagrammatic language.
Before we can do this, we must look at its original form and compare with the ZX rules discussed above.
To do this, we take the route presented in~\cite{eastSpinnetworksZXcalculus2022} and direct the reader seeking deeper explanation to~\cite{majorSpinNetwork1999}.

In the binor calculus, the identity over $\mathbb{C}^2$ is a single strand
\begin{equation}
\vcenter{\hbox{
    \includegraphics[height=1cm]{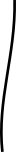}
}}
\cong
\dyad{0} + \dyad{1}.
\end{equation}
The free legs carry implicit labels of copies of $\mathbb{C}^2$.
There is also a duality between up and down.
These are given by the `cups' and `caps'.
The cup stands for
\begin{equation}
	\begin{tikzpicture}
	\begin{pgfonlayer}{nodelayer}
		\node [style=none] (0) at (-0.75, 0.75) {};
		\node [style=none] (1) at (0.75, 0.75) {};
	\end{pgfonlayer}
	\begin{pgfonlayer}{edgelayer}
		\draw [bend right=80, looseness=3.75] (0.center) to (1.center);
	\end{pgfonlayer}
\end{tikzpicture}
 \cong i \bra{01} - i \bra{10}.
\end{equation}
The cap is
\begin{equation}
	\begin{tikzpicture}
	\begin{pgfonlayer}{nodelayer}
		\node [style=none] (0) at (-0.75, -0.75) {};
		\node [style=none] (1) at (0.75, -0.75) {};
	\end{pgfonlayer}
	\begin{pgfonlayer}{edgelayer}
		\draw [bend left=80, looseness=3.75] (0.center) to (1.center);
	\end{pgfonlayer}
\end{tikzpicture}
 \cong i \ket{01} - i \ket{10}.
\end{equation}
With these definitions, the diagrams are then well-behaved under deformation: 
\begin{equation}
	\begin{array}{c}
		\includegraphics[height=1cm]{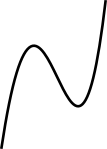}
	\end{array} = \begin{array}{c}
		\includegraphics[height=1cm]{figures/binor/straight}
	\end{array}
\end{equation}
Finally, the crossing is the regular swap, but with a global minus sign:
\begin{equation}
	\label{eq:cross_minus}
	\begin{array}{c}
		\includegraphics[height=1cm]{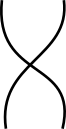}
	\end{array} = - \dyad{00} - \dyad{10}{01} - \dyad{01}{10} - \dyad{11}.
\end{equation}
These rules guarantee planar isotopy, i.e.\@ diagrams can be continuously deformed while preserving their interpretation as linear maps.

The cap and cup as Penrose diagrams can be translated into ZX diagrams as follows:
\[
	\begin{tikzpicture}
	\begin{pgfonlayer}{nodelayer}
		\node [style=none] (0) at (-0.75, -0.75) {};
		\node [style=none] (1) at (0.75, -0.75) {};
	\end{pgfonlayer}
	\begin{pgfonlayer}{edgelayer}
		\draw [bend left=80, looseness=3.75] (0.center) to (1.center);
	\end{pgfonlayer}
\end{tikzpicture}
\ \longmapsto\ \input{figures/binor/captranslation.tikz}
  \qquad \qquad \qquad
	\begin{tikzpicture}
	\begin{pgfonlayer}{nodelayer}
		\node [style=none] (0) at (-0.75, 0.75) {};
		\node [style=none] (1) at (0.75, 0.75) {};
	\end{pgfonlayer}
	\begin{pgfonlayer}{edgelayer}
		\draw [bend right=80, looseness=3.75] (0.center) to (1.center);
	\end{pgfonlayer}
\end{tikzpicture}
\ \longmapsto\ \input{figures/binor/cuptranslation.tikz}
\]
where for convenience, we ignore the imaginary number $i$ during the translation.
Similarly, though more importantly, the swap Penrose diagram is translated as the normal ZX swap without introducing the minus sign - below this has implication for how one structures the projectors to higher irreps.

\paragraph{Fundamental equations}

Binor calculus has two core equations that can be deduced from the definitions above.
\begin{equation}
	\label{eq:binor_circle}
	\begin{array}{c}
		\vcenter{\hbox{
    \includegraphics[height=1cm]{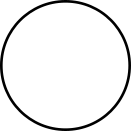}}} 
	\end{array}
	= -2
\end{equation}
\begin{equation}
	\label{eq:skein_relation}
	\begin{array}{c}
		\vcenter{\hbox{
    \includegraphics[height=1cm]{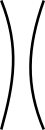}}}
	\end{array} + 
	\begin{array}{c}
		\vcenter{\hbox{
    \includegraphics[height=1cm]{figures/binor/cross}}}
	\end{array} + 
	\begin{array}{c}
		\vcenter{\hbox{
    \includegraphics[height=1cm]{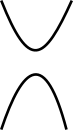}}}
	\end{array} = 0
\end{equation}
The value of the loop in the first equation secretly gives the \enquote{dimension} of the tensor calculus.
Here it is $-2$: the \enquote{2} explains the name \enquote{binor}, and the minus sign explains the title of the original article, \emph{Negative Dimensional Tensors} by Penrose~\cite{penroseApplicationsNegativeDimensional1971}.
The second equation is known under the name of \enquote{skein relation} or \enquote{binor identity}.
A glance at the ZX generators shows a difference between \cref{eq:binor_circle} and our Spin-ZX calculus is the minus sign.
The \enquote{skein relation}, however, does not exist in the same form (though of course still holds) as it relies on the negative sign incurred by the swap.
This difference rears its head more seriously when we construct the symmetric projector below.

\paragraph{(Anti-)symmetriser}
A key difference between the ZX formalism and Penrose's original diagrams is in how spaces are symmetrised.
In ZX we have seen the operator presented above in \cref{symmetriser} however the same mapping in Penrose's original diagrams to get the symmetrised space of multiple spin-$\frac{1}{2}$'s to give spin-$j$ irrep is given by anti-symmetrising the strands in binor calculus:
\begin{equation}\label{eq:Penrose-symmetrisation}
	\input{figures/binor/sym-j.tikz} 
	\overset{\text{def}}= \frac{1}{(2j)!}
	\sum_{\sigma \in \mathfrak{S}_{2j}} (-1)^{|\sigma|} \,  \input{figures/binor/sym-sigma.tikz} 
\end{equation}
where $|\sigma|$ is the parity of $\sigma$ and the $\sigma$-labelled box represents the corresponding permutation of the $2j$ strands from one side to the other.
Although this looks like an anti-symmetrisation (since we have the $(-1)^{|\sigma|}$ term), the operator is actually a projector from $\mathbb{C}^2 \otimes \cdots \otimes \mathbb{C}^2$ to $\mathcal{S}_{2j}(\mathbb{C}^2 \otimes \cdots \otimes \mathbb{C}^2 )$, because of the minus sign in \cref{eq:cross_minus}.
So we have
\begin{equation}\label{eq:binor-jm-vector}
	\input{figures/binor/sym-vec-binor2.tikz} = 
	\mathcal{S}_{2j} (\underbrace{\ket{0}...\ket{1}}_{j+m \text{ times } 0}) = \frac{1}{\sqrt{\binom{2j}{j-m}}} \ket{j;m}.
\end{equation}

Taking the trace of the symmetrised space (i.e.\@ making a loop) gives the value
\begin{equation}
	\input{figures/binor/circle.tikz} = 
	\  (-1)^{2j} (2j+1),
\end{equation}
which is the dimension of the space up to a phase.

\paragraph{Invariant tensors.}
Now, let us draw in binor calculus, the analogue of the 3-valent vertex, i.e.\@ the $3jm$-symbol.
The essential property to be noticed is the invariance of the cup under the action of $SU(2)$, i.e.\@ for any $u \in SU(2)$:
\begin{equation}
	\input{figures/binor/g-on-cups-2col.tikz}
\end{equation}
Then the diagram
\begin{equation}
	\input{figures/binor/3-valent-binor.tikz}
\end{equation}
depicts a vector that belongs to $\text{Inv}_{SU(2)} ( \mathcal{H}_{j_1} \otimes \mathcal{H}_{j_2} \otimes \mathcal{H}_{j_3})$, so it is proportional to $\ket{j_1,j_2,j_3}$.
This should be interpreted as a kind of \enquote{railroad switch}, where the fundamental wires within the three symmetrised bundles redistribute between themselves.
Because we are dealing with symmetrised spaces, we only care about how many wires go from each bundle to the other bundle.
It turns out that there is then actually only one way in which to connect the wires, when it is not impossible.
The Clebsch-Gordan conditions~\eqref{eq:CGcondition} precisely state when such a recoupling is possible.

Plugging the vectors of \cref{eq:binor-jm-vector}, one gets
\begin{widetext}
\begin{multline}
	\label{eq:penrose-3jm}
	\input{figures/binor/3-valent-binor-equal-3jm2.tikz} \\
	=
	N(j_1,j_2,j_3) i^{j_1+j_2+j_3} (-1)^{j_1-j_2+j_3}
	\begin{pmatrix}
		j_1 & j_2 & j_3 \\
		m_1 & m_2 & m_3
	\end{pmatrix}
	\frac{1}{\sqrt{\binom{2j_1}{j_1 - m_1}\binom{2j_2}{j_2 - m_2}\binom{2j_3}{j_3 - m_3}}}
\end{multline}
where
\begin{equation}\label{eq:N-factor}
	N(j_1,j_2,j_3) = \sqrt{\frac{(j_1+j_2+j_3+1)!(-j_1+j_2+j_3)!(j_1-j_2+j_3)!(j_1+j_2-j_3)!}{(2j_1)!(2j_2)!(2j_3)!}}.
\end{equation}
\end{widetext}

All the invariant functions that are introduced in the previous section can now be translated to binor diagrams.
We can then use the two core \cref{eq:binor_circle,eq:skein_relation}, which are sufficient to simplify the diagrams and compute their actual value.
We will often refer to the value $N(j_1,j_2,j_3)$ as the binor coefficient.

\section{Yutsis diagrams}\label{sec:yutsis}

The recoupling theory of $SU(2)$ can be graphically represented by Yutsis diagrams. There exist various conventions for Yutsis diagrams correspondence to recoupling in the literature. Here we use the original one, introduced by Yutsis in 1960~\cite{yutsisMathematicalApparatusTheory1962} and our explanation is taken from~\cite{eastPhysiqueFormelleSpin2022}.

\paragraph{3-valent node.}

The basic object of for Yutsis diagrams is the $3$-valent node, that represents Wigner's $3jm$ symbol:

\label{def:graphical 3jm}
\begin{align}\label{eq:graphical 3jm}
	\Wthree{j_1}{j_2}{j_3}{m_1}{m_2}{m_3} &= \begin{array}{c}
		\begin{overpic}[width=2.5cm]{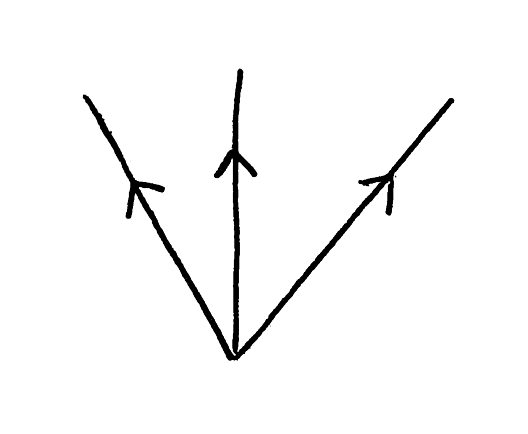}
			\put (-3,67) {$j_1m_1$}
			\put (37,73) {$j_2m_2$}
			\put (81,70) {$j_3m_3$}
			\put (45,5) {$-$}
		\end{overpic}
	\end{array} \nonumber \\
	&= \begin{array}{c}
		\begin{overpic}[width=2.5cm]{figures/yutsis/3CG-out.png}
			\put (-3,67) {$j_1m_1$}
			\put (37,73) {$j_3m_3$}
			\put (81,70) {$j_2m_2$}
			\put (45,5) {$+$}
		\end{overpic}
	\end{array}.
\end{align}

The signs $+/-$ on the nodes indicate the sense of rotation (anticlockwise/clockwise) in which the spins must be read. To alleviate notations we choose the default orientation to be minus.

Only the topology of the diagram matters, which means that all topological deformations are allowed.
\begin{align}
	\begin{array}{c}
		\begin{overpic}[width=2.5cm]{figures/yutsis/3CG-out.png}
			\put (-3,67) {$j_1m_1$}
			\put (37,73) {$j_2m_2$}
			\put (81,70) {$j_3m_3$}
		\end{overpic}
	\end{array} 
	&= \begin{array}{c}
		\begin{overpic}[width=2.5cm]{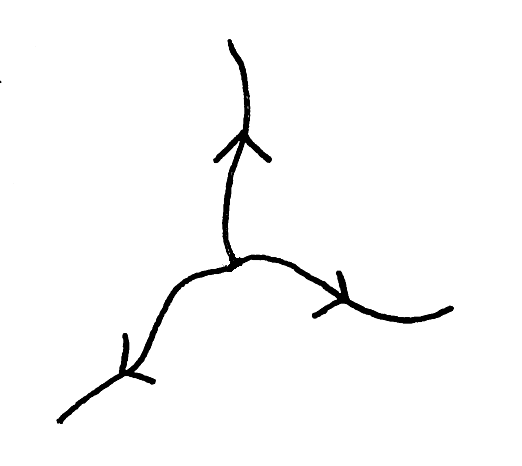}
			\put (0,-5) {$j_1m_1$}
			\put (35,85) {$j_2m_2$}
			\put (90,25) {$j_3m_3$}
		\end{overpic}
	\end{array} \nonumber \\
	&= \begin{array}{c}
		\begin{overpic}[width=2.5cm]{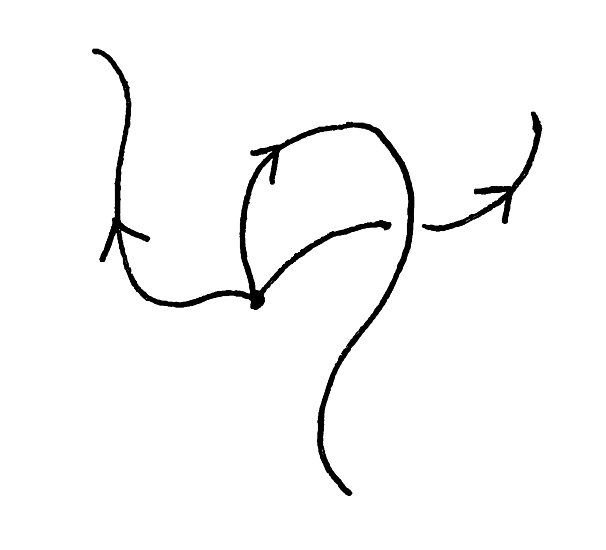}
			\put (5,85) {$j_1m_1$}
			\put (55,-5) {$j_2m_2$}
			\put (79,77) {$j_3m_3$}
		\end{overpic}
	\end{array}
\end{align}

For the arrows on the wires the \textit{ingoing} orientation corresponds to \textit{negating} the magnetic index. For instance
\begin{equation}\label{eq:3jm-wire inversion}
	\begin{array}{c}
		\begin{overpic}[width=2.5cm]{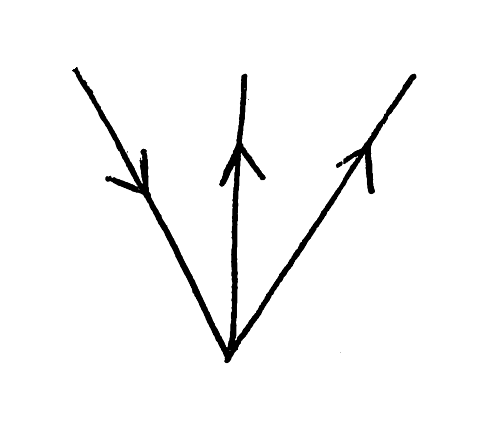}
			\put (-3,72) {$j_1m_1$}
			\put (37,74) {$j_2m_2$}
			\put (80,74) {$j_3m_3$}
		\end{overpic}
	\end{array}= \Wthree{j_1}{j_2}{j_3}{-m_1}{m_2}{m_3}\,.
\end{equation}

An important case is when one of the strands has spin 0. The spin-0 strand is then represented with a dashed line (no arrow needed):
\begin{equation}
	\begin{array}{c}
		\input{figures/yutsis/3-valent-0-strand.tikz}
	\end{array}
	= \frac{(-1)^{j_1+m_1}}{\sqrt{2j_1+1}}  \delta_{m_1,-m_3} \delta_{j_1,j_3}.
\end{equation}

We can then graphically define the two basic operations of multiplication and summation. Multiplication is implemented simply by juxtaposition of diagrams:
\begin{align}
	&\begin{array}{c}
		\begin{overpic}[width=2.5cm]{figures/yutsis/3CG-out.png}
			\put (-3,67) {$j_1m_1$}
			\put (37,73) {$j_2m_2$}
			\put (80,70) {$j_3m_3$}
		\end{overpic}
	\end{array} \quad \begin{array}{c}
		\begin{overpic}[width=2.5cm]{figures/yutsis/3CG-out.png}
			\put (-3,67) {$j_4m_4$}
			\put (37,73) {$j_5m_5$}
			\put (80,70) {$j_6m_6$}
		\end{overpic}
	\end{array} \nonumber \\
	= &\Wthree{j_1}{j_2}{j_3}{m_1}{m_2}{m_3} \Wthree{j_4}{j_5}{j_6}{m_4}{m_5}{m_6} 
\end{align}
Then, the gluing of two external wires with the same label $jm$, but opposite directions, defines the sum over $m$ (from $-j$ to $j$), with the additional factor $(-1)^{j-m}$ in the summand, like:
\begin{align}\label{eq:def_summation}
	&\begin{array}{c}
		\begin{overpic}[width=3cm]{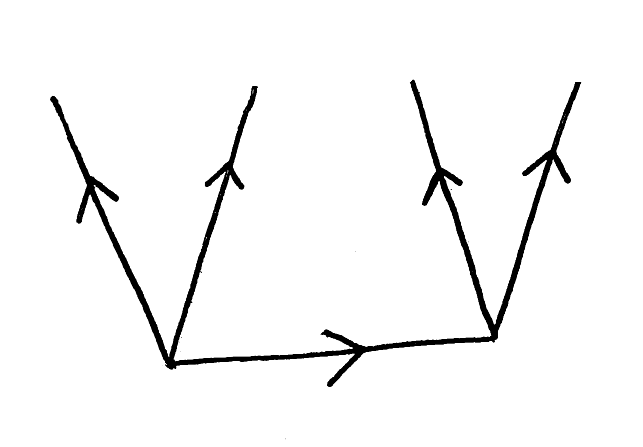}
			\put (-7,59) {$j_1m_1$}
			\put (25,59) {$j_2m_2$}
			\put (59,60) {$j_3m_3$}
			\put (91,60) {$j_4m_4$}
			\put (60,5) {$j$}
		\end{overpic}
	\end{array} \nonumber \\
	=
	&\sum_{m=-j}^j (-1)^{j-m}
	\begin{array}{c}
		\begin{overpic}[width=2.5cm]{figures/yutsis/3CG-out.png}
			\put (-3,70) {$j_1m_1$}
			\put (35,75) {$j_2m_2$}
			\put (85,68) {$jm$}
		\end{overpic}
	\end{array}
	\begin{array}{c}
		\begin{overpic}[width=2.5cm]{figures/yutsis/3CG-outin.png}
			\put (5,75) {$jm$}
			\put (37,73) {$j_3m_3$}
			\put (84,74) {$j_4m_4$}
		\end{overpic}
	\end{array}
\end{align}
On the RHS, we recognise the definition of the $4jm$-symbol:
\begin{equation}
	\begin{array}{c}
		\begin{overpic}[width=3cm]{figures/yutsis/4CG.png}
			\put (-7,59) {$j_1m_1$}
			\put (25,59) {$j_2m_2$}
			\put (59,60) {$j_3m_3$}
			\put (91,60) {$j_4m_4$}
			\put (60,5) {$j$}
		\end{overpic}
	\end{array} = 
	\Wfour{j_1}{j_2}{j_3}{j_4}{m_1}{m_2}{m_3}{m_4}{j} 
	\label{eq:CG4}
\end{equation}
The wire between the two nodes, whose magnetic index is summed over, is called an \textit{internal} wire. Reversing the arrow of an internal wire gives an overall phase:
\begin{equation}
	\begin{array}{c}
		\begin{overpic}[width=3cm]{figures/yutsis/4CG.png}
			\put (-10,59) {$j_1m_1$}
			\put (25,59) {$j_2m_2$}
			\put (59,60) {$j_3m_3$}
			\put (91,60) {$j_4m_4$}
			\put (60,5) {$j$}
		\end{overpic}
	\end{array}
	=(-1)^{2j}
	\begin{array}{c}
		\begin{overpic}[width=3cm]{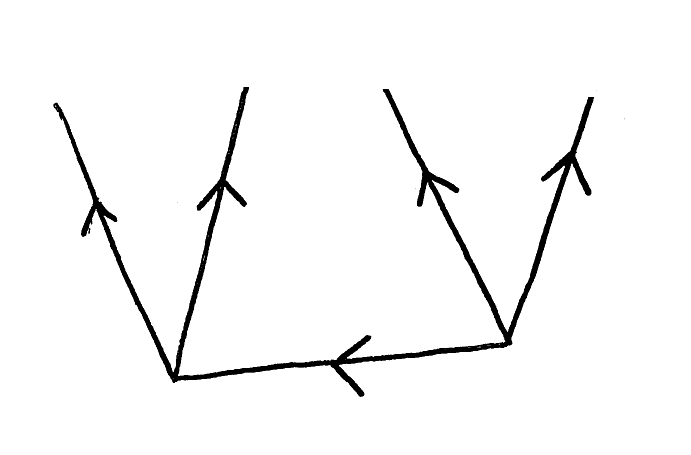}
			\put (-12,59) {$j_1m_1$}
			\put (20,59) {$j_2m_2$}
			\put (54,60) {$j_3m_3$}
			\put (86,60) {$j_4m_4$}
			\put (60,5) {$j$}
		\end{overpic}
	\end{array}.
	\nonumber
\end{equation} 
For convenience it is also helpful to consider a single wire with an arrow on it as an element of the graphical calculus:
\begin{align}
	\label{eq:yutsis_simple-strand}
	&\begin{array}{c}
		\begin{overpic}[width=1cm]{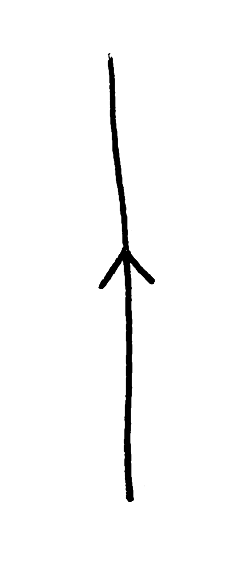}
			\put (15,95) {$j_1m_1$}
			\put (15,2) {$j_2m_2$}
		\end{overpic}
	\end{array}=\  (-1)^{j_2-m_2} \delta_{j_1j_2} \delta_{m_1m_2} & \qquad \qquad \nonumber \\
	\text{or} \qquad \qquad & & \qquad \qquad \nonumber \\
	&\begin{array}{c}
		\begin{overpic}[width=1cm]{figures/yutsis/up.png}
			\put (10,60) {$j$}
			\put (15,95) {$m$}
			\put (20,5) {$n$}
		\end{overpic}
	\end{array}= \ (-1)^{j-n} \delta_{mn} \,.& \qquad \qquad 
\end{align}
\refstepcounter{equation}

\section{Some General Lemmas}

\begin{lemma}
  \label{hopfditlm}\cite{wangQufiniteZXcalculusUnified2022}
  \begin{gather}
    \input{lemmas/hopfdit.tikz}
    \tag{Hopf}\label{rule:Hopf}\refstepcounter{equation}
  \end{gather}
\end{lemma}
\begin{lemma}
  \label{xdualiserlm}
  \begin{align*}
    &\input{lemmas/xdualiser-1-1.tikz}\\
    &\input{lemmas/xdualiser-1-2.tikz}\\
    &\input{lemmas/xdualiser-2-1.tikz}\\
    &\input{lemmas/xdualiser-2-2.tikz}
  \end{align*}
\end{lemma}

\hexagontwodaklt*
\begin{proof}
  We start from the Penrose-style honeycomb, where each site is the symmetric (spin-$\tfrac{3}{2}$) state of three qubits, one for each of its three singlet bonds.
  We rewrite each site into its Yutsis form using the isometry of \cref{prop:isometry-def}.
  Fusing and simplifying the resulting spiders and dualisers with the rules (S1), (S4), (D1) and (PK) labelled below then brings the lattice into the hexagonal layout.
  \[
    \scalebox{0.75}{\input{figures/qufit-applications/2daklt-proof-1.tikz}}
  \]
  \[
    \scalebox{0.75}{\input{figures/qufit-applications/2daklt-proof-2.tikz}}
  \]
  \[
    \scalebox{0.75}{\input{figures/qufit-applications/2daklt-proof-3.tikz}}
  \]
  \[
    \scalebox{0.75}{\input{figures/qufit-applications/2daklt-proof-4.tikz}}
  \]
  \[
    \scalebox{0.75}{\input{figures/qufit-applications/2daklt-proof-5.tikz}}
  \]
  On the last line, the dotted line indicates how the nodes are being rearranged in the last step.
\end{proof}

\section{Diagrammatic Proofs of Symmetriser Properties}\label{sec:symmetriser-proofs}
This section provides proofs of different properties of the symmetriser.
Note that some proofs are stated as equalities between Yutsis diagrams, while the proofs are given as a derivation between their ZX representations.
In such cases, we assume an implicit mapping from the given Yutsis diagrams to their ZX representations.

\isometrydef*
\begin{proof}
  We prove this by induction.
  \[
    \input{figures/binor/isometry-def-proof-2col.tikz}
  \]
\end{proof}

\begin{lemma}
  \[
    \input{figures/qufit-applications/symextend.tikz}
  \]
  where
  $m \geq n+1$,
  \begin{equation}
    \overrightarrow{a_{n+}} =
    \left(
      \frac{1}{\binom{n}{1}},  \cdots, \frac{1}{\binom{n}{n}},  \underbrace{x_1,\dotsc, x_s}_{m - 1-n}
    \right),
      \tag{SPN}\label{rule:Spn}\refstepcounter{equation}
  \end{equation}
  and $x_i$ are arbitrary complex numbers.
\end{lemma}
\begin{proof}
It follows from the Rules (DA) and (S1).
\end{proof}
\begin{lemma}\label{bialgebramix}
  \[
    \input{figures/qufit-applications/bialgebramixd2.tikz}
  \]
  where $s=\min\{x, m, n\}$, $t=\min\{y, m, n\}$. 
\end{lemma}
\begin{proof}
It follows from the Rules (S1), (DZX) and (B2).
\end{proof}

To show that $\mathcal{S}_n$ is a projector, we need the following lemmas:

\begin{lemma}
  \label{merge1}
  \[
    \input{figures/qufit-applications/supplementarydit.tikz}
  \]
  where $d\geq 2$, $n < d$, $\vec{b}=(\binom{n}{1},  \cdots, \binom{n}{k}, \cdots, \binom{n}{n}, \underbrace{0,\dotsc, 0}_{d - 1-n} )$.
\end{lemma}
\begin{proof}
  It follows from the rule (PA).
\end{proof}

\begin{lemma}
  \label{merge2}
  \[
    \input{figures/qufit-applications/supplementaryditfull.tikz}
  \]
\end{lemma}
\begin{proof}
  \[
    \input{figures/qufit-applications/supplementaryditfullprf.tikz}
  \]
\end{proof}

Now, we can show that the symmetriser is a projector:.
\symmetriseridep*
\begin{proof}
  \[
    \input{figures/qufit-applications/symmetriseridep-2col.tikz}
  \]
\end{proof}

The next property we show is \emph{invariance}, that is, that any $2 \times 2$ matrix commutes with the symmetriser.
To do this, we first derive some lemmas:

\begin{lemma}
  \label{zboxpushlm}
  \[
    \input{figures/qufit-applications/zboxpush.tikz}  
  \]
  where  $\vec{a}^n=(a,  \cdots, a^{k}, \cdots, a^{n}, \underbrace{0,\dotsc, 0}_{d - 1-n} )$, $d\geq 2$, $n < d$.
\end{lemma}
\begin{proof}
  It follows from the rule (PC).
\end{proof}

\begin{lemma}
  \label{1kplus1removerule}
  \[
  \input{figures/qufit-applications/1kplus1remove.tikz}
  \]
\end{lemma}
\begin{proof}
  It follows from the rule (PC).
\end{proof}

\begin{lemma}\label{zboxcommutelm}
  The qubit Z spiders could commute with the symmetriser:
  \[
    \input{figures/qufit-applications/zboxcommute.tikz}
  \]
\end{lemma}
\begin{proof}
  \[
    \input{figures/qufit-applications/zboxcommuteprf-2col.tikz}
  \]
\end{proof}
\begin{lemma}\label{hadamardcommutelm}
  The qubit Hadamard gates could commute with the symmetriser:
  \[
    \input{figures/qufit-applications/hadamardcommute.tikz}
  \]
\end{lemma}
\begin{proof}
  Since the qubit Hadamard gate is self-adjoint, it is equivalent to prove the following equality:
  \[
    \input{figures/qufit-applications/hadamardcommuteprf-2col.tikz}
  \]
  Therefore,
  \begin{align*}
    \input{figures/qufit-applications/hadamardcommuteprf2-2col-1.tikz}\\
    \input{figures/qufit-applications/hadamardcommuteprf2-2col-2.tikz}\\
    \input{figures/qufit-applications/hadamardcommuteprf2-2col-3.tikz}    
  \end{align*}
  The last equality holds because
  \[
    \input{figures/qufit-applications/hadamardcommuteprf3-2col.tikz}
  \]
  which is a power of $-1$.
\end{proof}

\invariance*
\begin{proof}
  Note that any $2 \times 2$ unitary can be decomposed as a product of Z spiders and Hadamard gates. So it follows directly from \cref{zboxcommutelm,hadamardcommutelm}. 
\end{proof}

\stacking*
\begin{proof}
  Let
  \begin{align*}
    \overrightarrow{a_{k+}} &= \left(
    \frac{1}{\binom{k}{1}}, \cdots, \frac{1}{\binom{k}{k}}, \underbrace{0,\dotsc, 0}_{n - k}
    \right),\\
    \overrightarrow{\frac{1}{a_{k+}}} &= \left(
    \binom{k}{1}, \cdots, \binom{k}{k}, \underbrace{0,\dotsc, 0}_{n - k}
    \right).
  \end{align*}
  Then
  \[
    \scalebox{.9}{\input{figures/qufit-applications/stackingprf-2col.tikz}}
  \]
\end{proof}

\cappingcupping*
\begin{proof}
  The first equality holds due to the following:
  \[
    \input{figures/qufit-applications/cappingprf-2col.tikz}
  \]
  The second equality can be proved similarly.
\end{proof}

\symmetrisersliding*
\begin{proof}
  \[
    \input{figures/qufit-applications/slidingprf-2col.tikz}
  \]
  where for the second equality we used the self-transpose property of the symmetriser.
\end{proof}

To prove the looping relation property, we need the following lemma:
\begin{lemma}
  \label{zboexsxmergelm}
  \input{figures/qufit-applications/zboexsxmerge.tikz}
\end{lemma}
where $\overrightarrow{\tau}=\left(
\frac{1}{\binom{n-1}{1}}, \cdots, \frac{1}{\binom{n-1}{n-1}}, \frac{2n}{n+1} \right).$
\begin{proof}
  It follows from the rule (PA).
\end{proof}

\loopingrelation*
\begin{proof}
  \[
    \input{figures/qufit-applications/loopingprf-2col.tikz}
  \]
\end{proof}

\section{Angular momentum proofs}

\jpjmcommute*
\begin{proof}
 First, since
 \begin{align*}
   [J_+, J_-]\ =\ J_+J_- - J_- J_+ \ &=\quad \input{figures/joperator/j+j-commute-1.tikz}\\
   &=\quad \input{figures/joperator/j+j-commute-2.tikz}
 \end{align*}
 and
 \[
   \input{figures/joperator/j+j-commute2-2col.tikz}
 \]
 We have
 \[
   [J_+, J_-]\ =\  \begin{tikzpicture}
	\begin{pgfonlayer}{nodelayer}
		\node [style=none] (227) at (0, -1) {};
		\node [style=gbox] (228) at (0, 0.025) {$A_j$};
		\node [style=none] (229) at (0, 1) {};
		\node [style=none] (231) at (-1, 0) {$2j$};
	\end{pgfonlayer}
	\begin{pgfonlayer}{edgelayer}
		\draw (228) to (227.center);
		\draw (228) to (229.center);
	\end{pgfonlayer}
\end{tikzpicture}
\ =\ 2J_3
 \]
 where we used
 \[
   \input{figures/joperator/j+j-commute4.tikz}
   \qquad \text{and} \qquad
   \input{figures/joperator/j+j-commute5.tikz}.
 \]
\end{proof}

\jthreejpcommute*
\begin{proof}
  \begin{align*}
    &[J_3, J_+]\ =\ J_3J_+ - J_+ J_3
    \ =\ \input{figures/joperator/j3j+commute-1-1.tikz}\hfill \\
    &=\ \input{figures/joperator/j3j+commute-1-2.tikz} \\
    &=\ \input{figures/joperator/j3j+commute-2-1.tikz} \\
    &=\ \input{figures/joperator/j3j+commute-2-2.tikz}
  \end{align*}
  where $A^{\prime}_j = \left(
    \frac{j-2}{j-1}, \frac{j-3}{j-1}, \cdots, \frac{j-2j}{j-1}, \frac{j}{j-1}
    \right), \quad A^{\prime\prime}_j = \frac{1}{j} \left(
    1, 1, \cdots, 1, -2j
    \right)$.
\end{proof}
\jthreejmcommute*
\begin{proof}
  \begin{align*}
    &[J_3, J_-]\ =\ J_3J_- - J_- J_3
    \ =\ \input{figures/joperator/j3j-commute-1-1.tikz}\hfill \\
    &=\ \input{figures/joperator/j3j-commute-1-2.tikz} \\
    &=\ \input{figures/joperator/j3j-commute-2-1.tikz} \\
    &=\ \input{figures/joperator/j3j-commute-2-2.tikz}
  \end{align*}
\end{proof}

\section{Diagrammatic Proofs of the Wigner Matrix}
\label{sec:wigmatrix}
\ifdefined\wignermatrixelement
\wignermatrixelement*
\fi
\begin{proof}
 \begin{align*}
 &\input{figures/binor/wignermatrixelemprf1-1.tikz}\\
 &\input{figures/binor/wignermatrixelemprf1-2.tikz}\\
 =& \frac{1}{\sqrt{\binom{2j}{j-n}}} \frac{1}{\sqrt{\binom{2j}{j-m}}}\\
 & \times \sum_{k} \frac{ (2j)! }{k!(j- m- k)!(j+n -k)!(m-n+k)!}\\
 & \times a^{j+n -k}b^{m-n+k}c^kd^{j-m -k} \\
 =& \sum_{k} \frac{ \sqrt{(j-m)!(j+m)!(j-n)!(j+n)!} }{k!(j- m- k)!(j+n -k)!(m-n+k)!}\\
 & \times a^{j+n -k}b^{m-n+k}c^kd^{j-m -k}
 \end{align*}
where $\begin{tikzpicture}
	\begin{pgfonlayer}{nodelayer}
		\node [style={rn_phase}] (64) at (0, 1) {$i_s\pi$};
		\node [style={rn_phase}] (73) at (0, -1.125) {$l_s\pi$};
		\node [style=box] (76) at (0, -0.075) {$u$};
	\end{pgfonlayer}
	\begin{pgfonlayer}{edgelayer}
		\draw (64) to (73);
	\end{pgfonlayer}
\end{tikzpicture}
\in \{a, b, c, d\}$.
 If the number of the term of $\begin{tikzpicture}
	\begin{pgfonlayer}{nodelayer}
		\node [style=rn] (64) at (0, 1) {};
		\node [style={rn_phase}] (73) at (0, -1.125) {$\pi$};
		\node [style=box] (76) at (0, -0.075) {$u$};
	\end{pgfonlayer}
	\begin{pgfonlayer}{edgelayer}
		\draw (64) to (73);
	\end{pgfonlayer}
\end{tikzpicture}
 = c $  in \input{figures/binor/isuls4terms.tikz}  is $k$, then the number of the term of $\begin{tikzpicture}
	\begin{pgfonlayer}{nodelayer}
		\node [style={rn_phase}] (64) at (0, 1) {$\pi$};
		\node [style={rn_phase}] (73) at (0, -1.125) {$\pi$};
		\node [style=box] (76) at (0, -0.075) {$u$};
	\end{pgfonlayer}
	\begin{pgfonlayer}{edgelayer}
		\draw (64) to (73);
	\end{pgfonlayer}
\end{tikzpicture}
 = d $ is $j-m-k$,  the number of the term of $\begin{tikzpicture}
	\begin{pgfonlayer}{nodelayer}
		\node [style={rn_phase}] (64) at (0, 1) {$\pi$};
		\node [style=rn] (73) at (0, -1.125) {};
		\node [style=box] (76) at (0, -0.075) {$u$};
	\end{pgfonlayer}
	\begin{pgfonlayer}{edgelayer}
		\draw (64) to (73);
	\end{pgfonlayer}
\end{tikzpicture}
 = b $ is $j-n-(j-m-k)=m-n+k$, and the number of the term of $\begin{tikzpicture}
	\begin{pgfonlayer}{nodelayer}
		\node [style=rn] (64) at (0, 1) {};
		\node [style=rn] (73) at (0, -1.125) {};
		\node [style=box] (76) at (0, -0.075) {$u$};
	\end{pgfonlayer}
	\begin{pgfonlayer}{edgelayer}
		\draw (64) to (73);
	\end{pgfonlayer}
\end{tikzpicture}
 = a $ is $j+n-k$.
 Therefore, by the formula of full permutations with repeated elements,  the number of the term of
$a^{j+n -k}b^{m-n+k}c^kd^{j-m -k}$ is $\frac{ (2j)! }{k!(j- m- k)!(j+n -k)!(m-n+k)!}$.
\end{proof}

\section{Diagrammatic Proofs of the 3jm-Symbol Properties.}\label{app:3jsymbols}
Irreducables are the fundamental building blocks from which other representations are built.
Any finite-dimensional representation of $SU(2)$ is completely reducible, meaning it can be written as a direct sum of irreps.
Specifically, a tensor product of irreps can be decomposed into a direct sum of irreps via a bijective intertwiner (or equivariant map).
Recoupling theory aims to study the space of such intertwiners.
\paragraph{Intertwiners.}
Given two representations $V$ and $W$ of a group $G$, a linear map $\iota : V \longrightarrow W$ is an \emph{intertwiner} when it commutes with the group action, meaning that
\begin{equation}
	\iota (g \cdot v) = g \cdot \iota (v)
\end{equation}
for all $g\in G$ and $v\in V$. 
It's easy to see that the set of intertwiners forms a vector space. We denote this vector space by $\text{Hom}_G(V,W)$.
This space $\text{Hom}_G(V,W)$ is isomorphic to $\text{Inv}_G(V \otimes W^*)$, where $W^*$ is the dual vector space of $W$ with the dual representation ($\rho^*(g) = \rho(g^{-1})^{\text{T}}$), and we define the \emph{invariant subspace} of any representation on $\mathcal{H}$ as
\begin{equation}
	\text{Inv}_G(\mathcal{H}) \ \overset{\text{def}}=\  \left\{\, \psi \in \mathcal{H} \mid \forall g \in G,\, g \cdot \psi = \psi \,\right\}.
\end{equation}
Equivalently, we can characterise it by
\begin{equation}
	\text{Inv}_G(\mathcal{H}) \ =\  \left\{\, \psi \in \mathcal{H} \mid \forall a \in \mathfrak{g},\, a \cdot \psi = 0 \,\right\}.
\end{equation}
Physicists use the term ``intertwiner'' for any vector in $\text{Inv}_G(\mathcal{H})$.

\paragraph{Spin networks.}
Given a collection of intertwiners, there is a way to weave these together into networks of spins for which the connective topology is all that matters.
\begin{definition}
  A \emph{spin network} is a triple $(\Gamma,L_\Gamma,\iota)$ that consists of
  \begin{itemize}
    \item a directed~\footnote{Though in the literature the directedness is often suppressed.} labelled graph $\Gamma$,
    \item a labelling $L_\Gamma$ that assigns an irrep (i.e.\@ spin) $j$ of $SU(2)$ to each edge $e$ of $\Gamma$, and
    \item a set of intertwiners $\iota$ that assigns to each vertex $v$ a mapping:
      \begin{equation}
        \ket{\iota_v} \in \operatorname{Inv}(\mathrm{v}, \mathrm{j}) \stackrel{\text { def }}{=} \operatorname{Inv}_{\mathrm{SU}(2)}\left(\bigotimes_{e \in \mathcal{S}(v)} H_{\mathfrak{j}_e} \otimes \bigotimes_{e \in \mathcal{T}(v)} H_{\mathfrak{j}_e}^\dagger \right).
      \end{equation}
    This is an $SU(2)$ equivariant~\footnote{That is, it lives in the space of $SU(2)$ invariant objects.} map from the irrep Hilbert spaces $H_{\mathfrak{j}_e}$ inbound source edges $\mathcal{S}(v)$ to the outbound target edge $\mathcal{T}(v)$ Hilbert spaces $H_{\mathfrak{j}_e}^\dagger$.
  \end{itemize}
\end{definition}

\subsection{3j state}
\begin{lemma}
  \label{kducom}
  \[
    \input{figures/3jsymbolcalc/kducomt.tikz}
  \]
  where $\overrightarrow{u} = \left(
        (-1)^1\binom{d}{1}, \cdots, (-1)^{d}\binom{d}{d}
      \right).$
\end{lemma}
\begin{widetext}
\begin{proof}
  \[
    \input{figures/3jsymbolcalc/kducomtprf.tikz}
  \]
 where $\overrightarrow{v} =  (-1)^{-d}\left(
        (-1)^0\binom{d}{0}, (-1)^1\binom{d}{1},\cdots, (-1)^{d-1}\binom{d}{d-1}
      \right).$ 
    
       \[ 
       \begin{split}
       \overrightarrow{v^{\prime}} =  (-1)^{-d}\left(
      (-1)^{d-1}\binom{d}{d-1},  (-1)^{d-2}\binom{d}{d-2},\cdots,     (-1)^0\binom{d}{0}   \right) \\
      = \left(
      (-1)^{-1}\binom{d}{1},  (-1)^{-2}\binom{d}{2},\cdots,     (-1)^{-d}\binom{d}{d}   \right) \\
      =  \left(
        (-1)^1\binom{d}{1}, \cdots, (-1)^{d}\binom{d}{d}
      \right) = \overrightarrow{u}.
      \end{split}
      \]
\end{proof}
\end{widetext}

Now we can derive the diagram which represent the state $\ket{j_1,j_2,j_3}$ as defined in \cref{eq:definition jstate}.
For this, we will need to use the following lemma.
\begin{lemma}\label{threejmaplem}
  For all pairs $(k,l) \in \{(1,2), (1,3), (2,3)\}$, we have:
  \[
    \input{3jsymbolcalc/simp-lemma-1.tikz}
  \]
  where $x_{12}=j_1+j_2-j_3$, $x_{13}=j_1+j_3-j_2$, $x_{23}=j_2+j_3-j_1$.
\end{lemma}
\begin{proof}
    \begin{align}
        &\input{3jsymbolcalc/simp-lemma-1-pf-1v2-2col.tikz}\nonumber\\
        &\input{3jsymbolcalc/simp-lemma-1-pf-2v2-2col.tikz}\nonumber\\
        &\input{3jsymbolcalc/simp-lemma-1-pf-3v2-2col.tikz}\label{eq:simp-lemma-1-pf}
    \end{align}
    where
    \begin{align*}
    &n_k = 2 j_k + 1,\\
    &\overrightarrow{\frac{1}{a^{-}_{x_{k \ell}}}}
      =
      \left(\!
         (-1)^1\binom{\!x_{k \ell}\!}{1},\!\cdots\!,(-1)^{x_{k \ell}}\binom{x_{k \ell}}{x_{k \ell}},
        \underbrace{0,\!\cdots\!, 0}_{n_k n_\ell - 1 - x_{k \ell}}
      \right)\!,\\
    &\text{and}\\
    &\overrightarrow{\frac{1}{a^{-}}}
      =
      \left(\!
        (-1)^1\binom{\!x_{k \ell}\!}{1},\!\cdots\!, (-1)^{x_{k \ell}}\binom{x_{k \ell}}{x_{k \ell}}
      \right).
    \end{align*}
\end{proof}
\ifdefined\statethenwignermatrices
\statethenwignermatrices*
\fi
\begin{proof}
  \[
    \scalebox{.8}{\input{figures/binor/g-on-cupszx-r-then-wigner-matrix-2col.tikz}}
  \]
\end{proof}

\begin{lemma}\label{threejmapoldnew}

  \[
    \input{3jsymbolcalc/3joldtonew-2col.tikz}
  \]
 
\end{lemma}

\begin{proof}
  \begin{align}
        &\input{figures/3jsymbolcalc/3jstatev3-4v3.tikz}\\
        &= \quad\input{figures/3jsymbolcalc/3jstatev3-5v2.tikz}\\
        &= \quad\input{figures/3jsymbolcalc/3jstatev3-6v2.tikz}\\
        &= \quad\input{figures/3jsymbolcalc/3jstatev3-7v2.tikz}
    \end{align}

\end{proof}

Then we can derive the diagram for the 3j state as follows:
\ifdefined\threejstate
\threejstate*
\fi
\begin{proof}
By definition, $\ket{j_1,j_2,j_3} \overset{\ref{3jstatedef}}{=}$
    \begin{align*}
       &\frac{(-1)^{j_1-j_2+j_3}}{N(j_1,j_2,j_3)} \tikzfigscale{0.9}{figures/binor/3-valent-binorzx}\\
       &\overset{\ref{threejmaplem}}{=} \quad \frac{(-1)^{x_{13}}}{N(j_1,j_2,j_3)}\tikzfigscale{0.9}{figures/3jsymbolcalc/3jstatev3-2v2}\\
       &\overset{\ref{threejmaplem}}{=} \quad\tikzfigscale{0.9}{figures/3jsymbolcalc/3jstatev3-4v2}\\
       &\overset{\ref{threejmapoldnew}}{=} \quad \frac{1}{N(j_1,j_2,j_3)} \tikzfigscale{0.9}{figures/3jsymbolcalc/3jstatev3-7v2}
    \end{align*}
\end{proof}

\ifdefined\threejsymbol
\threejsymbol*
\fi
\begin{proof}
  Below, we calculate the explicit value for the 3jm-symbol.
  \begin{align*}
  \left(
    \begin{array}{ccc}
      j_1 & j_2 & j_3 \\
      m_1 & m_2 & m_3
    \end{array}
    \right)
    \hspace{-3cm}&\\
      &= \quad\input{figures/3jsymbolcalc/3jm-symbolv2.tikz}\\
    &\overset{\ref{threejmapoldnew}}{=}  \quad\input{figures/3jsymbolcalc/3jmexplicit-alt-1-1.tikz}\\
    &\input{figures/3jsymbolcalc/3jmexplicit-alt-1-2.tikz}\\
    &=\input{figures/3jsymbolcalc/3jmexplicit-alt-2-1.tikz}\\
    &\input{figures/3jsymbolcalc/3jmexplicit-alt-2-2.tikz}\\
    &=\input{figures/3jsymbolcalc/3jmexplicit-alt-2plus-1.tikz}\\
    &\input{figures/3jsymbolcalc/3jmexplicit-alt-2plus-2.tikz}\\
    &= \input{figures/3jsymbolcalc/3jmexplicit-alt-3v2-1.tikz}\\
    &\input{figures/3jsymbolcalc/3jmexplicit-alt-3v2-2.tikz}
  \end{align*}
where
\begin{align*}
  C\ &=\ \frac{1}{\sqrt{\binom{2j_1}{j_1-m_1}}}\frac{1}{\sqrt{\binom{2j_2}{j_2-m_2}}}\frac{1}{\sqrt{\binom{2j_3}{j_3-m_3}}}\frac{1}{N(j_1,j_2,j_3)}.
\end{align*}

Now we need to know the range of the values $j_1 - m_1 - k$ and $j_2 - m_2 - x_{12} + k$. 

1)  Bounds for $j_1 - m_1 - k$.

i) First, we give a upper bound for $j_1 - m_1 - k$:
\begin{align*}
\begin{array}{c}
        -j_1 \leq m_1 \leq j_1 \Rightarrow     -j_1- m_1 \leq 0  \\
          k \geq 0  \Rightarrow     -j_1- m_1 \leq k  \\
 -j_1- m_1 \leq k  \Leftrightarrow          j_1-m_1-k \leq 2j_1
  \end{array} 
\end{align*}
Therefore, $ j_1-m_1-k \leq 2j_1 $.

ii) Then we give a lower bound for $j_1 - m_1 - k$ from the two constraints below, where the bottom of the two is the Clebsch-Gordan condition:
\begin{align*}
\left. \begin{array}{c}
     0 \leq k \leq x_{12}= j_1+j_2-j_3 \\
        j_1+j_2-j_3 \leq 2j_1 \Leftrightarrow  j_2-j_3 \leq j_1 
  \end{array} \right\}\\
  \Rightarrow k\leq 2j_1
\end{align*}
Since $m_1 \leq j_1$, we have $ k\leq 2j_1+ j_1-m_1$, which is equivalent to  $ j_1-m_1-k \geq -2j_1$. But this lower bound is not tight enough. To get a tighter one, we assume that  $j_1-m_1-k < 0$. Then we have
\begin{align*}
\left. \begin{array}{c}
     j_1-m_1-k \geq -2j_1 \\
      j_1-m_1-k < 0
  \end{array} \right\}\\
  \Rightarrow 
 1 \leq 2j_1+1+j_1-m_1-k < 2j_1+1
\end{align*}
 Furthermore,
\begin{align*}
\left. \begin{array}{c}
   k \leq  j_1+j_2-j_3 \\
   j_1-m_1 \geq 0
  \end{array} \right\}\\
  \Rightarrow 2j_1+1+j_1-m_1-k-x_{13}\\
  = 2j_1+1+j_1-m_1-k-(j_1+j_3-j_2)\\
  = j_1+j_2-j_3-k+j_1-m_1+1>0,
\end{align*}
 i.e.,
\[
  2j_1+1+j_1-m_1-k > x_{13}.
\]

Hence
\[
\input{figures/3jsymbolcalc/Kj1m1k.tikz}
\]
which means the value of the 3jm-symbol is $0$. To avoid this trivial case under the assumption $j_1-m_1-k < 0$, we can naturally set the lower bound as  $j_1-m_1-k \geq 0$.


2)  Bounds for $j_2-m_2-x_{12}+k $.
Substitute $x_{12}$ with its explicit value, we get 
\begin{align*}
  j_2-m_2-x_{12}+k\\
  =j_2-m_2-( j_1+j_2-j_3)+k\\
  =j_3-j_1-m_2+k.
\end{align*}

i) First we give a upper bound for $j_3-j_1-m_2+k$:
\begin{align*}
  \left. \begin{array}{c}
   k \leq  j_1+j_2-j_3 \\
   j_2+m_1 \geq 0
  \end{array} \right\}  \Rightarrow 
   k \leq  j_1+j_2-j_3+j_2+m_2
\end{align*}
which is equivalent to the following upper bound:
\[
j_3-j_1-m_2+k \leq 2j_2.
\]
ii) Second, we give a lower bound for $j_3-j_1-m_2+k$:
\begin{align*}
  \left. \begin{array}{c}
     k \geq 0 \\
    j_2+j_3-j_1\geq 0 \\
    j_2-m_2 \geq 0 
  \end{array} \right\} \Rightarrow 
  k \geq -( j_2+j_3-j_1+ j_2-m_2 )
\end{align*}
which is equivalent to the following upper bound:
\[
j_3-j_1-m_2+k \geq -2j_2.
\]
This lower bound is not tight enough. To get a tighter one, we assume that  $j_3-j_1-m_2+k < 0$. Then we have
\begin{align*}
\left. \begin{array}{c}
  j_3-j_1-m_2+k \geq -2j_2 \\
    j_3-j_1-m_2+k < 0
  \end{array} \right\}\\
  \Rightarrow 1 \leq 2j_2+1+j_3-j_1-m_2+k< 2j_2+1
 \end{align*}
 Moreover,
\begin{align*}
\left. \begin{array}{c}
   k \geq 0 \\
  j_2-m_2 \geq 0
  \end{array} \right\}\\
  \Rightarrow 2j_2+1+j_3-j_1-m_2+k-x_{23}\\
  = 2j_2+1+j_3-j_1-m_2+k-(j_2+j_3-j_1)\\
  = k+j_2-m_2+1> 0,
\end{align*}
  i.e.,
  \[
    2j_2+1+j_3-j_1-m_2+k > x_{23}.
\]
Therefore,
\[
\input{figures/3jsymbolcalc/Kj3j1m2k.tikz}
\]
which means the value of the 3jm-symbol is $0$. To avoid this trivial case under the assumption $j_3-j_1-m_2+k < 0$, we can naturally set the lower bound as  $j_3-j_1-m_2+k \geq 0$.


Combining the bounds we just obtained for $j_1 - m_1 - k$ and $j_2 - m_2 - x_{12} + k$ , we have
\[
  \left\{ \begin{array}{c}
            0 \leq k \leq x_{12}              \\
            0 \leq j_1-m_1-k \leq 2j_1        \\
            0 \leq j_2-m_2-x_{12}+k \leq 2j_2 \\
  \end{array}\right.
\]
To avoid the trivial case that the following diagram is 0,
\[
\input{figures/3jsymbolcalc/Kj3j1m2kKj1m1k.tikz}
\]
we also need
\[
  \left\{ \begin{array}{c}
            j_1-m_1-k \leq x_{13} \\
            j_2-m_2-x_{12}+k \leq x_{23}
  \end{array}\right.
\]
Since $x_{13}\leq 2j_1, x_{23} \leq 2j_2$, we then have
\begin{equation*}
  \left\{ \begin{array}{c}
    0 \leq k \leq x_{12}\\
    0 \leq j_1-m_1-k \leq x_{13}\\
    0 \leq j_2-m_2-x_{12}+k \leq x_{23}
  \end{array}\right.
\end{equation*}
which is equivalent to
\[
  \left\{ \begin{array}{c}
            k \geq 0           \\
            k \geq j_2-j_3-m_1 \\
            k \geq j_1-j_3+m_2 \\
            k \leq j_1-m_1     \\
            k \leq j_1+j_2-j_3 \\
            k \leq j_2+m_2
  \end{array}\right.
\]

Under these conditions, we have 
\begin{align*}
  &\left(
    \begin{array}{ccc}
      j_1 & j_2 & j_3 \\
      m_1 & m_2 & m_3
    \end{array}
  \right)\\
   &= \input{figures/3jsymbolcalc/3jmexplicit-alt-3v2plus-1.tikz}\\
   &\input{figures/3jsymbolcalc/3jmexplicit-alt-3v2plus-2.tikz}
  \end{align*}
  \begin{widetext}
  \begin{align*}
  &= C \sum_{k=K}^{N} (- 1)^{j_1 - m_1 + j_2 - m_2 - x_{12} + k+x_{13}} \binom{x_{12}}{k} \binom{x_{13}}{j_1 - m_1 - k} \binom{x_{23}}{j_2 - m_2 - x_{12} + k} \\
  &= \frac{1}{\sqrt{\binom{2j_1}{j_1-m_1}}}\frac{1}{\sqrt{\binom{2j_2}{j_2-m_2}}}\frac{1}{\sqrt{\binom{2j_3}{j_3-m_3}}}(-1)^{j_1-j_2+j_3} \\
  &\times N^{-1}(j_1,j_2,j_3)\sum_{k=K}^N (- 1)^{j_1 - m_1 + j_2 - m_2 - x_{12} + k} \binom{x_{12}}{k}\binom{x_{13}}{j_1\minu m_1 \minu k}\binom{x_{23}}{j_2\minu m_2\minu x_{12}\plus k} \\
  &= (-1)^{j_1-m_1+j_1-j_2+j_3+j_2\minu m_2\minu x_{12}}\frac{1}{\sqrt{\binom{2j_1}{j_1-m_1}}}\frac{1}{\sqrt{\binom{2j_2}{j_2-m_2}}}\frac{1}{\sqrt{\binom{2j_3}{j_3-m_3}}} \\
  &\times N^{-1}(j_1,j_2,j_3)\sum_{k=K}^N (\minu 1)^{k} \binom{x_{12}}{k}\binom{x_{13}}{j_1\minu m_1 \minu k}\binom{x_{23}}{j_2\minu m_2\minu x_{12}\plus k} 
\end{align*}
where
\[
  \begin{array}{l}
  D=  C \sum_{k=K}^{N} (- 1)^{j_1 - m_1 + j_2 - m_2 - x_{12} + k+x_{13}} \binom{x_{12}}{k} \binom{x_{13}}{j_1 - m_1 - k} \binom{x_{23}}{j_2 - m_2 - x_{12} + k},\\
    K= \max\{0, j_2-j_3-m_1,  j_1-j_3+m_2\}, \\
    N = \min\{ j_1-m_1, j_2+m_2, j_1+j_2-j_3\}, \\
   s= -(j_2 - m_2 - x_{12} + k)-(j_1 - m_1 - k)+ j_3+ m_3\\
    = -(j_2 - m_2 -(j_1+j_2-j_3)+k+j_1 - m_1 - k)+ j_3+ m_3 \\
    =-(j_3- m_2- m_1)+ j_3+ m_3 = m_2+m_1+ m_3 = 0.
  \end{array}
\]

We can further simplify by
\[
  \begin{array}{l}
  (-1)
    ^{j_1-m_1+j_1-j_2+j_3+j_2- m_2- x_{12}}= (-1)^{2j_1+j_3-m_1- m_2- (j_1+j_2-j_3)} \\
    = (-1)^{j_1-j_2+2j_3+m_3} =(-1)^{j_1+j_3-j_2} (-1)^{j_3+m_3}                     \\
    =(-1)^{j_1+j_3-j_2} (-1)^{-j_3-m_3} = (-1)^{j_1-j_2-m_3}.
  \end{array}
\]
Second,
\[
  \begin{array}{l}
    \frac{1}{\sqrt{\binom{2j_1}{j_1-m_1}}}\frac{1}{\sqrt{\binom{2j_2}{j_2-m_2}}}\frac{1}{\sqrt{\binom{2j_3}{j_3-m_3}}} = \sqrt{\frac{(j_1-m_1)!(j_1+m_1)!(j_2-m_2)!(j_2+m_2)!(j_3-m_3)!(j_3+m_3)!}{(2j_1)!(2j_2)!(2j_3)!}},\vspace{0.3cm}\\
    N(j_1,j_2,j_3)^{-1}=\sqrt{\frac{(2j_1)!(2j_2)!(2j_3)!}{(j_1+j_2+j_3+1)!(j_2+j_3-j_1)!(j_1+j_3-j_2)!(j_1+j_2-j_3)!}},\vspace{0.3cm}\\
    \binom{x_{12}}{k}\binom{x_{13}}{j_1\minu m_1 \minu k}\binom{x_{23}}{j_2\minu m_2\minu x_{12}\plus k}= \frac{x_{12}!}{k!(x_{12}-k)!}\frac{x_{13}!}{(j_1\minu m_1 \minu k)!(x_{13}-(j_1\minu m_1 \minu k))!}\frac{x_{23}!}{(j_2\minu m_2\minu x_{12}\plus k)!(x_{23}-(j_2\minu m_2\minu x_{12}\plus k))!}\vspace{0.3cm}\\
    =\frac{(j_2+j_3-j_1)!(j_1+j_3-j_2)!(j_1+j_2-j_3)!}{k!(j_1+j_2-j_3-k)!(j_1\minu m_1 \minu k)!(j_3-j_2+m_1 +k)!(j_3-j_1-m_2+k)!(j_2+m_2-k)!}
  \end{array}
\]

With these simplifications and binomial coefficients expansions, we have
\[
  \begin{array}{ll}
    \left( \begin{array}{ccc}
             j_1 & j_2 & j_3 \\
             m_1 & m_2 & m_3
    \end{array} \right) = (-1)^{j_1-j_2-m_3}\sqrt{\frac{(j_2+j_3-j_1)!(j_1+j_3-j_2)!(j_1+j_2-j_3)!}{(j_1+j_2+j_3+1)!}} & \vspace{0.3cm}\\
    \times \sqrt{(j_1-m_1)!(j_1+m_1)!(j_2-m_2)!(j_2+m_2)!(j_3-m_3)!(j_3+m_3)! } & \vspace{0.3cm} \\
    \times  \sum_{k=K}^N \frac{(-1)^{k}}{k!(j_1+j_2-j_3-k)!(j_1\minu m_1 \minu k)!(j_3-j_2+m_1 +k)!(j_3-j_1-m_2+k)!(j_2+m_2-k)!}
  \end{array}
\]
If we let the value of the 3j-symbol to be $0$ for the case when $m_1+m_2+m_3 \neq 0$, then we get
\[
  \begin{array}{ll}
    \left( \begin{array}{ccc}
             j_1 & j_2 & j_3 \\
             m_1 & m_2 & m_3
    \end{array} \right) = \delta(m_1+m_2+m_3, 0)(-1)^{j_1-j_2-m_3}\sqrt{\frac{(j_2+j_3-j_1)!(j_1+j_3-j_2)!(j_1+j_2-j_3)!}{(j_1+j_2+j_3+1)!}} & \vspace{0.3cm}\\
    \times \sqrt{(j_1-m_1)!(j_1+m_1)!(j_2-m_2)!(j_2+m_2)!(j_3-m_3)!(j_3+m_3)! } & \vspace{0.3cm} \\
    \times  \sum_{k=K}^N \frac{(-1)^{k}}{k!(j_1+j_2-j_3-k)!(j_1\minu m_1 \minu k)!(j_3-j_2+m_1 +k)!(j_3-j_1-m_2+k)!(j_2+m_2-k)!}
  \end{array}
\]
where $\delta(i, j)$ is the Kronecker delta, which is exactly the same formula as given in~\cite{shorePrinciplesAtomic1967}.  
\end{widetext}

\end{proof}

\subsection{Injection map}
\ifdefined\injectionmap
\injectionmap*
\fi
\begin{proof}
We first prove that the injection $\mathcal{H}_{j_3} \longrightarrow \mathcal{H}_{j_1} \otimes \mathcal{H}_{j_2} $ as given in (\ref{injectionmap}) can be represented by the following diagram:
 \[ \scalebox{0.75}{\input{figures/3jsymbolcalc/cgcoefficient.tikz}} \]
We just need to prove that if we plug $\ket{j_3-m_3}$ on top and $\bra{j_1-m_1}\otimes \bra{j_2-m_2}$ on bottom of the diagram,  which is $ C^{j_3m_3}_{j_1m_1j_2m_2}$ by definition, then we will get ~\cite{martin-dussaudPrimerGroupTheory2019}
\[
    C^{j_3m_3}_{j_1m_1j_2m_2}
    \quad=\quad
  (-1)^{-j_1 + j_2 - m_3} \sqrt{2 j_3 + 1}  \left(
    \begin{array}{ccc}
      j_1 & j_2 & j_3 \\
      m_1 & m_2 & -m_3
    \end{array}
  \right).
\]
In fact, we have
\begin{align*}
  &\scalebox{0.9}{\input{figures/3jsymbolcalc/cgcoefficientprf1-1.tikz}} \\
  =&\scalebox{0.9}{\input{figures/3jsymbolcalc/cgcoefficientprf1-2.tikz}} \\
  =&\scalebox{0.9}{\input{figures/3jsymbolcalc/cgcoefficientprf1-3.tikz}} \\
  =&\scalebox{0.9}{\input{figures/3jsymbolcalc/cgcoefficientprf2-1.tikz}} \\
  =&\scalebox{0.9}{\input{figures/3jsymbolcalc/cgcoefficientprf2-2.tikz}} \\
  =& (-1)^{-j_1 + j_2 - m_3} \sqrt{2 j_3 + 1}  \left(
    \begin{array}{ccc}
      j_1 & j_2 & j_3 \\
      m_1 & m_2 & -m_3
    \end{array}
  \right).
\end{align*}
where
\begin{align*}
 (-1)^{ j_3 - m_3}(-1)^{ x_{13}} &= (-1)^{ m_3 - j_3}(-1)^{ j_1 + j_3 - j_2}\\
 =(-1)^{ j_1-j_2 + m_3} &= (-1)^{-j_1 + j_2 - m_3}.
\end{align*}
Secondly, we have
  \begin{align*}
  \tikzfigscale{0.9}{figures/3jsymbolcalc/cgcoefficientsimplify-1}\\
  \tikzfigscale{0.9}{figures/3jsymbolcalc/cgcoefficientsimplify-2}\\
  \tikzfigscale{0.9}{figures/3jsymbolcalc/cgcoefficientsimplify-3}
  \end{align*}
\end{proof}

\subsection{Symmetry properties of 3j-symbols}
\ifdefined\threejsymbolsignflip
\threejsymbolsignflip*
\fi
\begin{proof}
  \begin{align*}
    &\phantom{=} \ \left(
      \begin{array}{ccc}
        j_1    & j_2    & j_3    \\
        \!-m_1 & \!-m_2 & \!-m_3
      \end{array}
    \right)\\
    &\overset{\ref{threejmapoldnew}}{=} \ \input{figures/3jsymbolcalc/3jmvaluepermute-alt-1.tikz}\\
    &\input{figures/3jsymbolcalc/3jmvaluepermute-alt-2-1.tikz}\\
    &\input{figures/3jsymbolcalc/3jmvaluepermute-alt-2-2.tikz}\\
    &\input{figures/3jsymbolcalc/3jmvaluepermute-alt-3-1.tikz}\\
    &\input{figures/3jsymbolcalc/3jmvaluepermute-alt-3-2.tikz}\\
    &\input{figures/3jsymbolcalc/3jmvaluepermute-alt-4-1.tikz}\\
    &\input{figures/3jsymbolcalc/3jmvaluepermute-alt-4-2.tikz}\\[10pt]
    &= \quad (-1)^{j_1 +j_2 + j_3}
    \left(
      \begin{array}{ccc}
        j_1 & j_2 & j_3 \\
        m_1 & m_2 & m_3
      \end{array}
    \right)
  \end{align*}
  where
  \begin{align*}
    C &= \frac{1}{\sqrt{\binom{2j_1}{j_1+m_1}}}\frac{1}{\sqrt{\binom{2j_2}{j_2+m_2}}}\frac{1}{\sqrt{\binom{2j_3}{j_3+m_3}}}\frac{1}{N(j_1,j_2,j_3)}\\
    &= \frac{1}{\sqrt{\binom{2j_1}{j_1-m_1}}}\frac{1}{\sqrt{\binom{2j_2}{j_2-m_2}}}\frac{1}{\sqrt{\binom{2j_3}{j_3-m_3}}}\frac{1}{N(j_1,j_2,j_3)}
  \end{align*}
\end{proof}

\section{Diagrammatic Calculations of Quantised Volume}\label{app:volume}

\volumeoperator*
\begin{proof}
  \[
    \scalebox{0.75}{\input{figures/qufit-applications/vol-state-1.tikz}}
  \]
  \[
    \scalebox{0.75}{\input{figures/qufit-applications/vol-state-2.tikz}}
  \]
  \[
    \scalebox{0.75}{\input{figures/qufit-applications/vol-state-3.tikz}}
  \]
  \[
    \scalebox{0.75}{\input{figures/qufit-applications/vol-state-4.tikz}}
  \]
  \[
    \scalebox{0.75}{\input{figures/qufit-applications/vol-state-5.tikz}}
  \]
  \[
    \scalebox{0.75}{\input{figures/qufit-applications/vol-state-6.tikz}}
  \]
  \[
    \scalebox{0.75}{\input{figures/qufit-applications/vol-state-7.tikz}}
  \]
  By applying the following rules,
  \[
    \scalebox{0.75}{\input{figures/qufit-applications/dimpiv2-2col.tikz}}
  \]
  we get
  \[
    \scalebox{0.75}{\input{figures/qufit-applications/vol-state-8.tikz}}
  \]
  \[
    \scalebox{0.75}{\input{figures/qufit-applications/vol-state-9.tikz}}
  \]
  \[
    \scalebox{0.75}{\input{figures/qufit-applications/vol-state-10.tikz}}
  \]
  \[
    \scalebox{0.75}{\input{figures/qufit-applications/vol-state-11.tikz}}
  \]
\end{proof}



\begin{acknowledgments}
We would like to thank Nathan Fitzpatrick, Tuomas Laakkonen, and Richie Yeung for their feedback on the paper and suggestions for improvement.
RS is supported by the Clarendon Fund Scholarship.
RDPE acknowledges financial support from a Blaumann Foundation grant and useful conversations with A. Grushin.
LY acknowledges funding from the Google PhD Fellowship and from the Tencent Dowling Postdoctoral Fellowship.
BP is supported by the Engineering and Physical Sciences Research Council grant number EP/Z002230/1, \enquote{(De)constructing quantum software (DeQS)}.
\end{acknowledgments}

\bibliography{references-bibtex.bib}

\end{document}